\documentclass[reprint,amsmath,amssymb, aps]{revtex4-2}

\usepackage{graphicx}
\usepackage{dcolumn}
\usepackage{bm}
\usepackage{tabularray}
\usepackage{amsmath,amssymb}
\usepackage{subcaption}
\usepackage{xcolor}
\begin{document}
\preprint{APS/123-QED}

\title{Competition between the neutron-proton pair break-ups delineating the level structure of $^{202}$Po}
\author{Sahab~Singh$^1$}
\email{sahab.19phz0002@iitrpr.ac.in}
\author{D.~Choudhury$^1$}%
\author{B.~Maheshwari$^{2}$}
\author{R.~Roy$^1$}%
\author{K.~Yadav$^1$}%
\author{R.~Palit$^3$}
\author{B.~Das$^3$}
\author{P.~Dey$^3$}
\author{A.~Kundu$^3$}
\author{Md.~S.~R.~Laskar$^3$}
\author{D.~Negi$^3$}
\author{V.~Malik$^3$}
\author{S.~Jadhav$^3$}
\author{B.~S.~Naidu$^3$}
\author{A.~V.~Thomas$^3$}
\author{D.~L.~Balabanski$^4$}
\author{A.~Dhal$^4$}
\author{S.~Bhattacharya$^5$}
\author{A.~K.~Singh$^6$}
\author{S.~Bhattacharyya$^7$}
\author{S.~Nag$^8$}
\affiliation{$^1$Department of Physics, Indian Institute of Technology Ropar, Rupnagar, Punjab-140001, INDIA}
\affiliation{$^1$Grand Accélérateur National d’Ions Lourds, CEA/DSM-CNRS/IN2P3, B. P. 55027, F-14076, Caen Cedex 5, France}
\affiliation{$^3$Tata Institute of Fundamental Research, Mumbai-400005, INDIA}
\affiliation{$^4$Extreme Light Infrastructure - Nuclear Physics, M\u{a}gurele-077125, ROMANIA}
\affiliation{$^5$Racah Institute of Physics, The Hebrew University of Jerusalem, Jerusalem-91904, ISRAEL}
\affiliation{$^6$Department of Physics, Indian Institute of Technology Kharagpur, West Bengal-721302, INDIA}
\affiliation{$^7$Variable Energy Cyclotron Centre, 1/AF Bidhannagar, Kolkata-700064, INDIA}
\affiliation{$^8$Department of Physics, Indian Institute of Technology BHU, Varanasi-221005, INDIA}

\date{\today}%

\begin{abstract}
High-spin spectroscopic study of $^{202}$Po ($Z$ = 84, $N$ = 118) has been carried out using the $^{195}$Pt($^{12}$C, 5n)$^{202}$Po fusion-evaporation reaction. An extended level scheme has been proposed up to an excitation energy of $E_x\approx$ 8 MeV and angular momentum of 27$\hbar$, with the addition of 57 newly observed $\gamma$-ray transitions, along with the revisions in the placement of 8 already known transitions and the multipolarities of 4 of these transitions. The energy of the unobserved 8$^+ \rightarrow 6^+$ transition has been proposed to be 9.0(5) keV, which resolves the uncertainty in the excitation energy of the levels above the 6$^{+}$ state. Three new sequences of $M1$ transitions have also been identified in the high excitation energy regime and included in the proposed level scheme. The large-scale shell model calculations for $Z>82$ and $N<126$ valence space have been carried out using PBPOP interaction which explained the overall level scheme for both the positive and negative parity states.
The calculations successfully reproduced the purity of the proton $\pi h_{9/2}$ dominated $8^+$ isomeric state, and also explained the missing $E2$ decay of the ${12}^+$ isomeric state in terms of changing nucleonic configurations.
\end{abstract}
\maketitle
\section{\label{sec:level1}Introduction}

Nuclei around the doubly-magic shell closure ($Z = 82$, $N = 126$) have significant physics interest as they exhibit both single-particle excitations at low spins, and a variety of collective rotational behavior at high spins~\cite{Janssens1991, Frauendorf}. 
These nuclei manifest a variety of interesting structural properties such as shape coexistence~\cite{Heyde1983,Wood1992}, shears bands~\cite{Clark2000}, neutron core excitations across $N$ = 126 shell closure, presence of a variety of short- and long-lived isomers~\cite{Atlas,Dracoulis2016}, and appearance of superdeformed bands~\cite{Janssens1991}. Nuclei lying in this transitional region, where level structures are governed due to competing single-particle and collective excitations, provide a fertile ground for the testing of large-scale shell-model calculations~\cite{McGrory1975,Caurier2003}.

The even-even lighter Po isotopes ($A = 200-208$) with two protons above the $Z = 82$ shell closure, typically have a competition between the two quasi-proton and two quasi-neutron configurations~\cite{Fant1990,Beuscher1976}. Several efforts have been made to understand the low-lying states in these isotopes~\cite{Bijnens1998,Hagemann1971}. Results of new measurements with in-source laser spectroscopy and multi-step Coulomb excitation hinted towards an early departure of Po isotopes from sphericity on moving away from the $N = 126$ shell closure, as compared to their neighboring $Z\leq82$ isotones~\cite{Cocolios2011,Kesteloot2015}. A transition in the ground state deformation of these nuclei has also been observed from spherical and near spherical to oblate and then prolate deformed structure close to $N = 104$ mid-shell.~\cite{Helariutta1999,Van2003}.
 
The existence of isomeric states is rather common in the nuclei in $A$ = 200 mass region due to competing neutron-proton configurations~\cite{Atlas, Dracoulis2016, Tandel2024}. In even-even Po isotopes, the yrast $8^+$ state has been classified as fairly pure $\pi(h_{9/2})^2$ state and found to be isomeric in nature, systematically~\cite{Hausser1976,Nagamiya1972,Maheshwari2021}.
The yrast $12^+$ isomeric state is usually understood in terms of $\nu(i_{13/2})^2$ configuration~\cite{Maj1990} de-exciting to the $10^+$ state by an $E2$ transition~\cite{Fant1990,Maj1990,HCJain1986,Baldsiefen2001}, except for the case of $N=118$, $^{202}$Po, where the $12^+$ isomeric state is known to de-excite by $E1$ transitions to the first and second excited $11^-$ states~\cite{Fant1990}. The energy systematics in the even-even $^{200-208}$Po isotopes~\cite{Fant1990} shows that the $11_2^-$ state has a higher excitation energy than the yrast $12^+$ isomeric state for $N<118$, while the ordering is reversed for $N\geq118$, leading to the possibility for the $12^+\rightarrow11^-_2$ de-excitations. Interestingly, no $10^+$ state has yet been observed in the $^{202}$Po, unlike the neighboring even-even Po isotopes, hinting towards different structural phenomena in this nucleus. Moreover, a theoretical study on seniority isomers in the Pb and Hg isotopes~\cite{Maheshwari2021} using the generalized seniority approach, predicted seniority to be a nearly good quantum number in the Po ($Z$ = 84, i.e. two-proton particle) isotopes as well and suggested $N = 118$ to be a transition point~\cite{Maheshwari2021} nucleus, which needs further investigations.

Moreover, several isomeric states have been reported in this region, which de-excite via enhanced $E3$ transitions~\cite{Dracoulis2016,Podoly2024}.
Such transitions mainly involve high-$j$ proton ($j_{15/2}$, $g_{9/2}$) and neutron ($i_{13/2}$, $f_{7/2}$) orbitals which differ by $\Delta j = \Delta l$ = 3. An $11^-$ isomeric state is systematically known in even-even $^{196–210}$Po~\cite{Maj1986,Wikstrom1974}. In the lighter Po isotopes ($N \leq$ 118), this isomer is found to decay via competing $E2$ and $E3$ transitions and an increasing trend in the $B(E3)$ values was noted~\cite{Maj1986}, indicating an increased collectively on moving away from the $N=126$ shell closure.

In addition, regular sequences of magnetic dipole transitions, first observed in the Pb isotopes in 1990s~\cite{Baldsiefen1992,Clark1992,Hubel1992,Jain2010}, have been identified in nuclei around $Z$ = 82, many of which were  identified as shears bands~\cite{Baldsiefen1994,Gorgen2001,Pai2014,Clark1993,Auranen2018,Kanjilal2022,Hartley2008} and understood through the tilted axis cranking and covariant density functional theory~\cite{Hubel1992,Jain2010,Frauendorf1993,Zhao2011,Meng2013,Meng2016,Yu2012}. No such M1 sequences are known in any of the lighter Po isotopes calling for new measurements to explore the high-spin structure of these nuclei. 

A complete spectroscopy of $^{202}$Po nucleus enables us to explore the role of the two protons above the shell closure giving rise to the various interesting aforementioned phenomena as well as digging into the uniqueness of the $N=118$ case. The work presented here was carried out with this motivation. Prior to this work, high-spin states in $^{202}$Po were investigated by Fant \textit{et al.}~\cite{Fant1990} up to $J^{\pi}$ = (22$^+$). Also, the low-medium spin states up to ($9^-$) were explored by Bijnens \textit{et al.}~\cite{Bijnens1998} via $\beta$-decay study of $^{202}$At. Recently, the half-lives of the $E2$-decaying $8^+$ and $E3$-decaying $11^-_1$ isomeric states were re-measured by our group~\cite{Sahab2024} using the electronic timing of the high-purity germanium (HPGe) clover detectors~\cite{Laskar2021}.
The present work reports the results from a detailed spectroscopic study of $^{202}$Po. An extended level scheme of $^{202}$Po is presented up to an angular momentum of 27$\hbar$ and excitation energy ($E_x)\sim$ 8 MeV, with the addition of 57 new $\gamma$-ray transitions. The experimental results were understood with large-scale shell-model calculations performed using PBPOP interaction.

\section{Experimental details}

The excited states of $^{202}$Po were populated and investigated utilizing the $^{195}$Pt($^{12}$C, 5n)$^{202}$Po fusion-evaporation reaction. The experiment was performed at Tata Institute of Fundamental Research (TIFR), Mumbai, where the 14-UD BARC-TIFR Pelletron LINAC facility provided the $^{12}$C beam at 83- and 87 MeV energies. The $^{12}$C beam was bombarded on a 3.2 mg/cm$^{2}$ thick, isotopically enriched (97.3\% enrichment) $^{195}$Pt target, with a $^{197}$Au catcher foil. The de-exciting $\gamma$ rays were detected using a hybrid array of 16 Compton-suppressed HPGe clover detectors strategically arranged in six rings, viz. $-23^{\circ}$, $\pm40^{\circ}$, $\pm65^{\circ}$, and $90^{\circ}$ with respect to the beam direction, coupled with 14 ($2'' \times 2''$) LaBr$_3$(Ce) scintillation detectors. The target-to-detector distance was 25~cm. 

A mixed source consisting of $^{152}$Eu and $^{133}$Ba isotopes, placed at the target position, was used to calibrate the $\gamma$-ray energy and efficiency of the HPGe clovers and LaBr$_3$(Ce) scintillators. A digital data acquisition (DDAQ) system based on Pixie-16 modules developed by XIA LLC~\cite{Palit2012,Tan2008} was used for collecting the in-beam data. Two- and higher-fold coincidence events were collected in list mode format. Two-crate synchronization method was used, with one crate for six digitizer modules with 100-MHz sampling frequency for the HPGe clovers, and the other for 1 module with 250-MHz sampling frequency for the LaBr$_3$(Ce) detectors. This paper reports the results of high-spin spectroscopy obtained using the clover HPGe detectors.
\section{Data Analysis}
\begin{figure*}[ht!]
\includegraphics*[height=1.05\textwidth,width=1.0\textwidth,angle=-90]{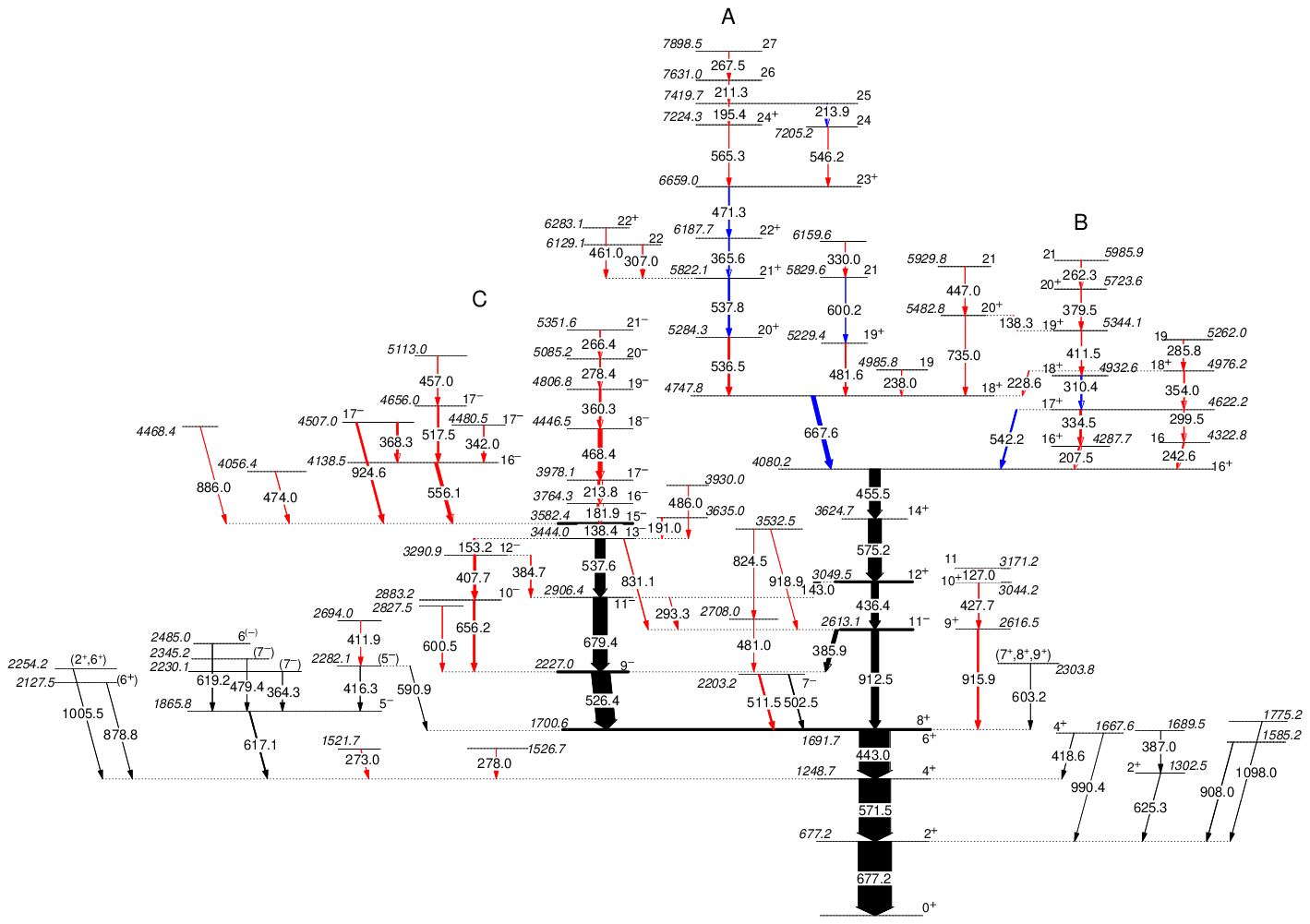}
\caption{\label{fig1:LS202po} Level scheme of $^{202}$Po obtained from the present work. The excited levels and the corresponding de-exciting $\gamma$ rays are labeled, with their energies in keV. The previously known $\gamma$-ray transitions confirmed in the present work are shown in black color, while those which are revised in the present work are shown in blue. The transitions newly observed in the present work are shown in red. The details of spin-parity assignments are given in Table~\ref{table:dco}, and elaborated in the text.}
\end{figure*}
%
Offline calibration and gain matching were carried out using the DAMM analysis program~\cite{DAMM}. The $\gamma\gamma$ coincidence events were sorted into $E_{\gamma}$-$E_{\gamma}$ symmetric and angle-dependent asymmetric matrices using the data sorting routine “Multi pARameter time-stamped based COincidence Search program (MARCOS)” developed at TIFR~\cite{Palit2012}. The $E_{\gamma}$-$E_{\gamma}$-$E_{\gamma}$ cubes were also constructed for coincidence analysis using the three and higher-fold coincidence events. A total of $\sim1.5*10^{10}$ double ${\gamma}$-ray events and $\sim6*10^{9}$ triple ${\gamma}$-ray events were recorded. The matrices and cubes were investigated with different coincidence windows of 100 and 300 ns, respectively. The Radware software package~\cite{Radford1995-1,Radford1995-2} was used for further data analysis. The level scheme of the $^{202}$Po nucleus was constructed considering the coincidence relationships between the observed $\gamma$ rays, their relative intensities, multipolarities and electromagnetic characters.

The spins of the levels were assigned on the basis of the multipole order of the de-exciting $\gamma$-ray transitions, which in turn were obtained from the measurements of Directional Correlation of $\gamma$-ray de-exciting Oriented states (DCO ratio method)~\cite{Kramer1989,Krane1973}. The DCO ratios were measured using an asymmetric matrix with the \textit{x}-axis and \textit{y}-axis corresponding to the detectors at 157$^{\circ}$ and 90$^{\circ}$, respectively. The DCO ratio,
\begin{eqnarray}
    R_{DCO}=\frac{I^{\gamma_1}_{157^{\circ}}(Gate^{\gamma_2}_{90^{\circ}})}{I^{\gamma_1}_{90^{\circ}}(Gate^{\gamma_2}_{157^{\circ}})}
    \label{DCO}
\end{eqnarray}
was found to be about 0.5(1.0) and 1.0(2.0) for pure dipole and quadrupole transitions, respectively, with the gate on stretched quadrupole (dipole) transitions. These values were determined from the transitions of $^{202}$Po with known multipolarities. Intermediate values of the $R_{DCO}$ ratio indicate a mixed nature of the transitions. The $R_{DCO}$ value for the unstretched dipole transitions have been observed to be similar to that of the quadrupole transitions. 

\begin{figure*}[ht!]
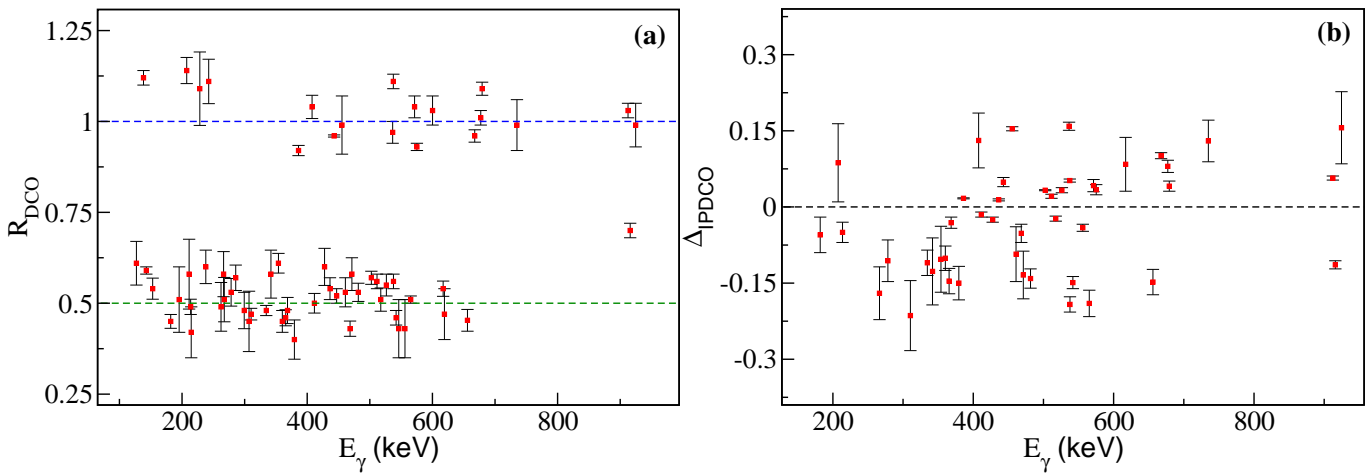

        \centering
        \begin{subfigure}[b]{0.50\textwidth}
                \centering
                \includegraphics*[width=\textwidth]{fig2a.eps}
                \label{fig1a:spectrum}
        \end{subfigure}%
        \begin{subfigure}[b]{0.508\textwidth}
                \centering
                \includegraphics*[trim=0 0
 0 0,clip,width=\textwidth]{fig2b.eps} 
                \label{fig1b:levelscheme}
        \end{subfigure}%
           \caption{(a) $R_{DCO}$ values for the observed $\gamma$-ray transitions, wherever possible, deduced using the gates of stretched quadrupole $\gamma$ rays. The green and blue dotted lines in the y-axis are the eye-guides for the DCO ratios depicting pure dipole and quadrupole transitions, respectively. (b) $\Delta_{IPDCO}$ values for various observed $\gamma$ rays. The dotted line in the y-axis is the eye-guide for $\Delta_{IPDCO}$ values differentiating the electric (positive $\Delta_{IPDCO}$) and magnetic (negative $\Delta_{IPDCO}$) nature of the transitions. 
           } 
        \label{fig:DCOPol}
\end{figure*}
The parity of the excited states were determined from the linear polarization measurement of the decaying $\gamma$-ray transitions. The approach of integrated polarization directional correlation from oriented states (IPDCO) was used to extract the polarization asymmetry~\cite{Starosta1999,Droste1996}. To identify the electric/magnetic character of the $\gamma$ rays, four clover detectors placed at 90$^{\circ}$ with respect to the beam direction, were used as Compton polarimeters~\cite{Jones1995}. The electric or magnetic character of the $\gamma$ rays was determined from the asymmetry of the Compton-scattered events in parallel and perpendicular directions.
The experimental polarization asymmetry, $\Delta_{IPDCO}$, was calculated as
\begin{eqnarray}
    \Delta_{IPDCO}=\frac{a(E_\gamma)N_{\perp}-N_{\parallel}}{a(E_\gamma)N_{\perp}+N_{\parallel}},
    \label{sqrt}
\end{eqnarray}
where, $N_{\perp}$($N_{\parallel}$) denotes the number of $\gamma$-ray counts scattered perpendicular (parallel) to the reaction plane. The geometrical asymmetry correction factor $a(E_{\gamma})$ of the detection setup, was determined using the Compton-scattered events of the $\gamma$ rays from unpolarized $^{152}$Eu and $^{133}$Ba radioactive sources, and is specified as $a(E_{\gamma})=N_{\parallel}/N_{\perp}$. By fitting the experimentally observed a($E_\gamma$) values at different energies with a linear expression, $a(E_\gamma) = a_0 + a_1E_\gamma$, we obtained $a_0 = +0.993(5)$ and $a_1 = -5.6(7) \times 10^{-5}$ keV$^{-1}$. This result indicated a negligible dependence of $a(E_\gamma)$ on the $\gamma$-ray energy ($E_\gamma$) over the energy range considered for the purpose.

In order to determine the polarization asymmetry ($\Delta_{IPDCO}$), two asymmetric matrices corresponding to the parallel and perpendicular segments of the clover detectors (with respect to the reaction plane) were constructed, where x-axis of both the matrices consisted of $\gamma$ rays from all other detectors and the y-axis for the two different matrices had $\gamma$ rays scattered in the parallel and perpendicular segments, respectively, of the clover detectors placed at $90^{\circ}$. A positive $\Delta_{IPDCO}$ indicates the stretched electric or non-stretched magnetic ($M1$) transition, while a negative value implies stretched magnetic or non-stretched electric ($E1$) transition.
%
\section{Experimental Results}\label{results}
Figure~\ref{fig1:LS202po} presents the level scheme of $^{202}$Po as obtained from the present work, where the newly observed $\gamma$-ray transitions are shown in red color. A total of 57 $\gamma$-ray transitions have been newly identified and added to the previously known level scheme~\cite{Fant1990,Bijnens1998} along with the spin-parity assignments of the new and already known levels, wherever possible. 
The level scheme up to the yrast 16$^{+}$ state is found to be consistent with that reported by Fant {\it et al.}~\cite{Fant1990}, except for the placement of a 180.7-keV transition as discussed in the following sections. The known transitions and energy levels above the yrast 16$^{+}$ state~\cite{Fant1990}, are revised and these transitions are shown with blue color in Fig.~\ref{fig1:LS202po}. The $\gamma$-ray energies, relative intensities, $R_{DCO}$ ratios, $\Delta_{IPDCO}$ values and the assigned multipolarities of the observed $\gamma$-ray transitions are listed in Table~\ref{table:dco}, along with the level energies ($E_x$) and spin-parity ($J^{\pi}$) assignments to the corresponding energy levels. The energy error is approximately 0.1 keV for the intense ($I_{\gamma}>20$) $\gamma$ rays and ranges between 0.2-0.5 keV for the lower-intensity $\gamma$ rays. The relative $\gamma$-ray intensities of the observed transitions have been measured in coincidence with different transitions using a symmetric $\gamma\gamma$ matrix and finally normalized to the intensity of the 677.2-keV transition for which the total intensity was assumed to be 100 units. The deduced $R_{DCO}$ and $\Delta_{IPDCO}$ values for the different $\gamma$-ray energies have been plotted in the figure.
\begin{table*}
         \caption{\label{table:dco}The $\gamma$-ray energies ($E_{\gamma}$), level energies ($E_x$), and the relative intensities ($I_{\gamma}$) of the $\gamma$-ray transitions are shown. The measured values of the $R_{DCO}$ ratios (with gates on stretched quadrupole transitions), $\Delta_{IPDCO}$ values, the proposed multipolarity of the $\gamma$ rays, and the spin-parities of the corresponding initial and final states ($J^{\pi}_i {\rightarrow}J^{\pi}_f$) are also listed. The quoted uncertainties in the $\gamma$-ray energies and the relative $\gamma$-ray intensities have the contribution from both the systematic and statistical errors. The systematic uncertainty in $I_\gamma$ is considered as 5$\%$ of $I_\gamma$.}
                  \centering
    \begin{ruledtabular}
        \begin{tabular}{ccccccc}
           $E_{\gamma}$(keV) & $E_x$(keV) & $I_{\gamma}$ & $R_{DCO}$ & $\Delta_{IPDCO}$ & Multipolarity &$J^{\pi}_i {\rightarrow}J^{\pi}_f$ \\
\hline      
           127.0(4) & 3171.2(8) & 1.5(2) & 0.61(6)\footnotemark[6] & & $D$& 11 $\rightarrow 10^+$ \\
           138.4(2) & 3582.4(5) & 11.8(7) & 1.12(2)\footnotemark[6] & & $E2$ & $15^-\footnotemark[1] \rightarrow 13^-$  \\
           138.3(4) & 5482.8(6) &  $<0.5$ &  & & & $20^+ \rightarrow 19^+$  \\
           143.0(2) & 3049.5(4) & 16.5(10) & 0.59(1)\footnotemark[6] & & $E1$ & $12^+ \rightarrow 11^-$  \\
           153.2(3) & 3444.0(4) & 1.4(2) & 0.53(3)\footnotemark[7] & & ${M1}$  & $13^- \rightarrow 12^-$ \\
           181.9(3)& 3764.3(6) & 5.9(6) & 0.45(2)\footnotemark[7] & -0.055(35) & $M1$ &$16^- \rightarrow 15^-$ \\
           191.0(5) & 3635.0(7) & 0.6(4) &  &  &  & $\rightarrow 13^-$\\
           195.4(4) & 7419.7(10) & 0.8(3) & 0.51(9)\footnotemark[8] & &  $D$ & $25 \rightarrow 24^+$ \\
           207.5(4) & 4287.7(5) & 3.7(3) & 1.14(4)\footnotemark[9] & 0.087(77) & $M1/E2$ & $16^+ \rightarrow 16^+$\\
           211.3(5) & 7631.0(11) & 0.7(3) & 0.58(9)\footnotemark[8] &  & $D$ & $26 \rightarrow 25 $\\
           213.8(4) & 3978.1(7) & 5.7(4) & 0.49(2)\footnotemark[7] & -0.050(20) & $M1$ & $17^- \rightarrow 16^-$\\
           213.9(5) & 7419.7(10)  & 0.8(2) & 0.42(7)\footnotemark[8] &  & $D$ & $25 \rightarrow 24$\\
           228.6(4) & 4976.2(5) & 0.8(2) & 1.08(10)\footnotemark[8] &  & $M1/E2$ & $18^+ \rightarrow 18^+$\\
           238.0(5) & 4985.8(7) & 0.7(1) & 0.60(4)\footnotemark[8] & & $D$ & $19 \rightarrow 18^+$\\
           242.6(4) & 4322.8(5) & 1.7(2) & 1.11(6)\footnotemark[9] &  & $D$ & $16 \rightarrow 16^+$\\
           262.3(5) & 5985.9(9) & 0.7(1) & 0.49(7)\footnotemark[9] &  & $D$ & $21 \rightarrow 20^+$\\
           266.4(4) & 5351.6(10) & 1.1(1) & 0.58(6)\footnotemark[7] & -0.170(52) & $M1$ & $21^- \rightarrow 20^-$\\
           267.5(4) & 7898.5(11) & 0.6(1) & 0.52(6)\footnotemark[8] &  & $D$ & $27 \rightarrow 26$\\
           273.0(4) & 1521.7(4) & 0.8(3) &  &  &  & $\rightarrow 4^+$\\
           278.0(5) & 1526.7(5) & $<0.5$  &  &  &  & $\rightarrow 4^+$\\
           278.4(4) & 5085.2(9) & 1.4(1) & 0.53(4)\footnotemark[7] & -0.106(41) & $M1$ & $20^- \rightarrow 19^-$\\
           285.8(5) & 5262.0(7) & 0.9(2) & 0.56(4)\footnotemark[9] &  & D & {$19 \rightarrow 18^+$}\\
           293.3(5) & 2906.4(4) & 0.6(1) &  &  &  &$11^- \rightarrow 11^-$\\
           299.5(5) & 4622.2(5) & 1.3(2) & 0.48(5)\footnotemark[9] &  & $D$  & $17^+ \rightarrow 16$\\
           307.0(4) & 6129.1 (8)& 1.1(2) & 0.45(8)\footnotemark[8] &  & $D$ & $22 \rightarrow 21^+$\\
           310.4(4) & 4932.6(6) & 2.8(3) & 0.47(2)\footnotemark[9] & -0.214(69)& $M1$ & $18^+ \rightarrow 17^+$\\
           330.0(5) & 6159.6(9) & $<0.5$  &  &  &  & $\rightarrow 21$\\
           334.5(4) & 4622.2(5) & 3.8(4) & 0.48(2)\footnotemark[9] & -0.110(25) & $M1$ & $17^+ \rightarrow 16^+$\\
           342.0(4) & 4480.5(6) & 1.2(2) & 0.58(7)\footnotemark[7] & -0.127(66) & $M1$ & $17^- \rightarrow 16^-$\\
           354.0(4) & 4976.2(5) & 1.8(5) & 0.61(3)\footnotemark[9] & -0.103(65) & $M1$ & $18^+ \rightarrow 17^+$\\
           360.3(4) & 4806.8(8) & 3.7(3) & 0.45(3)\footnotemark[7] & -0.101(24) & $M1$ & $19^- \rightarrow 18^-$\\
           364.3(4) & 2230.1(5) & $<0.5$ & & &  &$(7^-)$\footnotemark[2] $\rightarrow 5^-$\\
           365.6(4) & 6187.7(8) & 2.1(3) & 0.46(2)\footnotemark[8] & -0.146(25)& $M1$ & $22^+ \rightarrow 21^+$\\
           368.3(3) & 4507.0(5) & 4.3(4) & 0.48(4)\footnotemark[7] & -0.031(11) &  $M1$ & $17^- \rightarrow 16^-$\\
           379.5(5) & 5723.6(8) & 0.9(1) & 0.40(5)\footnotemark[9] & -0.150(33) & $M1$ & $20^+ \rightarrow 19^+$\\
           384.7(4) & 3290.9(5) & 1.1(1) &  &  & &$12^- \rightarrow 11^-$\\
           385.9(2) & 2613.1(4) & 10.4(9) & 0.92(2)\footnotemark[6] & 0.017(1) & $E2$  & $11^- \rightarrow 9^-$\\
           387.0(5) & 1689.5(7)& 0.5(5) &    &    &   & $\rightarrow 2^{+}\footnotemark[2]$ \\
           407.7(3) & 3290.9(5) & 6.1(5) & 1.03(3)\footnotemark[7] & 0.131(54) & $E2$ & $12^- \rightarrow 10^-$\\
           411.5(4) & 5344.1(6) & 1.5(3) & 0.50(3)\footnotemark[9] & -0.015(5) & $M1$ & $19^+ \rightarrow 18^+$\\
           411.9(5) & 2694.0(6) & $<0.5$ &  &  &  & $\rightarrow (5^{-}$)\footnotemark[2]\\
           416.3(4) & 2282.1(4) & 0.8(2) &  &  &  & $(5^-)\footnotemark[2] \rightarrow 5^-$\\         
           418.6(5) & 1667.6(5) & $<0.5$ &  &  & & $4^+\footnotemark[2] \rightarrow 4^+$\\
           427.7(4) & 3044.2(6) & 1.7(2) & 0.59(5)\footnotemark[6] & -0.025(5) & $M1$ & $10^+ \rightarrow 9^+$\\
           436.4(1) & 3049.5(4) & 20.8(12) & 0.54(3)\footnotemark[6] & 0.014(2)  & $E1$  & $12^+ \rightarrow 11^-$\\
           443.0(1) & 1691.7(2) & 91.0(48) & 0.96(3)\footnotemark[10] & 0.049(9) & $E2$ & $6^+\rightarrow 4^+$\\
           447.0(4) & 5929.8(7) & 1.7(1)  & 0.52(2)\footnotemark[8] &  & $D$ & $21\rightarrow 20^+$\\
           455.5(1) & 4080.2(4) & 31.2(17) & 0.98(8)\footnotemark[11] & 0.154(4) & $E2$ & $16^+ \rightarrow 14^+$\\
           457.0(4) & 5113.0(7) & 1.0(2) &  &  &  & $\rightarrow 17^-$\\
           461.0(4) & 6283.1(8) & 1.0(2) & 0.53(4)\footnotemark[8] & -0.093(54) & $M1$ & $22^+ \rightarrow 21^+$\\
           468.4(3) & 4446.5(7) & 8.2(13) & 0.43(2)\footnotemark[7] & -0.052(18) & $M1$ & $18^- \rightarrow 17^-$\\
           471.3(4) & 6659.0(9) & 1.9(3) & 0.58(5)\footnotemark[8] & -0.134(47) & $M1$ & $23^+ \rightarrow 22^+$\\
           474.0(4) & 4056.4(6) & 0.7(3) & & & & $\rightarrow 15^-$\\
           479.4(4) & 2345.2(5) & $<0.5$ &  & &  & $(7^{-})\footnotemark[2] \rightarrow 5^-$\\
           481.0(5) & 2708.0(6) & $<0.5$ &  &  &  & $\rightarrow 9^-$\\
           481.6(4) & 5229.4(8) & 2.0(2) & 0.53(3)\footnotemark[8] & -0.141(19) & $M1$ &$19^+ \rightarrow 18^+$\\                                         
        \end{tabular}
    \end{ruledtabular}
\end{table*}
\begin{table*}
         \centering
    \begin{ruledtabular}
        \begin{tabular}{ccccccc}
           $E_{\gamma}$(keV) & $E_x$(keV) & $I_{\gamma}$(rel.) & $R_{DCO}$ & $\Delta_{IPDCO}$ & Multipolarity & $J^{\pi}_i {\rightarrow}J^{\pi}_f$ \\
           \hline
           486.0(5) & 3930.0(7) & $<0.5$  &  &  &  & $\rightarrow 13^-$\\
           502.5(4) & 2203.2(6) & 2.4(2) & 0.57(3)\footnotemark[6] & 0.033(1) & $E1$ &$7^- \rightarrow 8^+$\\
           511.5(3) & 2203.2(6) & 4.5(9)\footnotemark[4] & 0.56(2)\footnotemark[6] & 0.021(4) & $E1$ & $7^- \rightarrow 6^+$\\
           517.5(3) & 4656.0(6) & 1.5(2) & 0.51(3)\footnotemark[7] & -0.023(5) & $M1$ &$17^- \rightarrow 16^-$\\
           526.4(1) & 2227.0(4) & 55.9(29) & 0.55(3)\footnotemark[6] & 0.033(5) & $E1$ &$9^- \rightarrow 8^+$\\
           536.5(3) & 5284.3(6) & 5.0(3) & 0.97(2)\footnotemark[8] & 0.159(8) & $E2$ &$20^+ \rightarrow 18^+$\\
           537.6(1) & 3444.0(4) & 26.1(16) & 1.11(2)\footnotemark[7] & 0.052(3) & $E2$ &$13^- \rightarrow 11^-$\\
           537.8(3) & 5822.1(6) & 3.6(2) & 0.56(2)\footnotemark[8] & -0.192(15) & $M1$ &$21^+ \rightarrow 20^+$\\
           542.2(3) & 4622.2(5) & 3.7(3) & 0.46(2)\footnotemark[9] & -0.149(12) & $M1$ &$17^+ \rightarrow 16^+$\\ 
           546.2(4) & 7205.2(9) & 0.8(1) & 0.43(8)\footnotemark[8] &  & $D$ &$24 \rightarrow 23^+$\\
            556.1(2) & 4138.5(5) & 7.6(4) & 0.43(8)\footnotemark[7] & -0.041(7) & $M1$ & $16^- \rightarrow 15^-$\\
            565.3(4) & 7224.3(9) & 1.2(2) & 0.51(1)\footnotemark[8] & -0.190(26) & $M1$ &$24^+ \rightarrow 23^+$\\
            571.5(1) & 1248.7(1) & 94.0(47) & 1.04(3)\footnotemark[6] & 0.042(12) & $E2$ &$4^+ \rightarrow 2^+$\\
            575.2(1) & 3624.7(4) & 39.0(13) & 0.93(1)\footnotemark[6] & 0.034(10) & $E2$ & $14^+ \rightarrow 12^+$\\
            590.9(5) & 2282.1(4) & $<0.5$  &  &  &  & ($5^{-})\footnotemark[2] \rightarrow 6^+$\\
            600.2(5) & 5829.6(5) & 0.7(1)  & 1.03(4)\footnotemark[8]  &  & $Q$ & $21 \rightarrow 19^+$\\
            600.5(5) & 2827.5(6) & 1.0(1) & & &  & $ \rightarrow 9^-$\\
            603.2(5) & 2303.8(6)&  $<0.5$ &   &  &  & $(7^+,8^+,9^+)\footnotemark[2] \rightarrow 8^+$\\
            617.1(3) & 1865.8(3) & 3.1(5) & 0.54(2)\footnotemark[10] & 0.084(53) & $E1$ &$5^- \rightarrow 4^+$\\
            619.2(4) & 2485.0(5) & 1.4(3) & 0.47(7)\footnotemark[10] & & $M1$ & $6^{(-)}\footnotemark[3] \rightarrow 5^-$\\
            625.3(5) & 1302.5(5) & $<0.5$ & & &  & $2^+$\footnotemark[2] $\rightarrow 2^+$\\
            656.2(3) & 2883.2(5) & 4.7(6) & 0.45(3)\footnotemark[7] & -0.148(25) & $M1$ & $10^- \rightarrow 9^-$\\
            667.6(2) & 4747.8(5) & 11.7(7) & 0.95(2)\footnotemark[9] & 0.101(6) & $E2$ & $18^+ \rightarrow 16^+$\\
            677.2(1) & 677.2(1) & 100(5) & 1.01(2)\footnotemark[10] & 0.080(12) & $E2$ & $2^+ \rightarrow 0^+$\\
            679.4(1) & 2906.4(4) & 42.2(21) & 1.09(2)\footnotemark[6] & 0.041(10) &  $E2$ & $11^- \rightarrow 9^-$\\
            735.0(4) & 5482.8(6) & 1.3(1) & 0.99(7)\footnotemark[8] & 0.130(41) & $E2$ & $20^+ \rightarrow 18^+$\\
            824.5(5) & 3532.5(6) & $<0.5$ &  &  &  & \\
            831.1(4) & 3444.0(4) & 1.5(1) &  &  & & $13^- \rightarrow 11^-$\\
            878.8(5) & 2127.5(5) & $<0.5$ &  &  &  & $(6^+)\footnotemark[2] \rightarrow 4^+$\\
            886.0(5) & 4468.4(7) & $<0.5$ &  &  &  &$\rightarrow 15^{-}$\\
            908.0(4) & 1585.2(4) & 0.8(2) &  &  & &$\rightarrow 2^+$ \\
            912.5(1) & 2613.1(4) & 21.5(13) & 1.03(2)\footnotemark[6] & 0.057(4) & $E3$ &$11^-\footnotemark[5] \rightarrow 8^+$\\
            915.9(3) & 2616.5(5) & 4.7(4) & 0.70(2)\footnotemark[6] & -0.114(8) & $M1+E2$ &$9^+ \rightarrow 8^+$\\
            918.9(5) & 3532.5(6) & $<0.5$ &  &  &  & $\rightarrow 11^-$\\
            924.6(3) & 4507.0(5) & 4.8(3) & 0.99(6)\footnotemark[7] & 0.156(71) & $E2$ &$17^- \rightarrow 15^-$\\
            990.4(5) & 1667.6(5) & 0.4(1)&  &  &  & $4^{+}\footnotemark[2] \rightarrow 2^+$ \\
            1005.5(5) & 2254.2(5) & $<0.5$ &  &  & & $(2^+,6^+)\footnotemark[2] \rightarrow 4^+$\\
            1098.0(5) & 1775.2(5) & $<0.5$ &  &  & & $\rightarrow 2^{+}$ \\
        \end{tabular}
    \end{ruledtabular}
    \footnotetext[1]{parity is adopted from Ref.~\cite{Fant1990}.}
    \footnotetext[2]{spin and parity are adopted from Ref.~\cite{Bijnens1998}, except for the 2345.2-keV level, for which $J^{\pi}$ = $(7^{-})$ was tentatively assigned with the assistance of the $E$2 multipolarity of the corresponding de-exciting 479.4-keV $\gamma$ ray as adopted from Ref.~\cite{Bijnens1998} and the shell-model calculations in the present work. See the text for the details.}
   \footnotetext[3]{Parity is adopted from Ref.~\cite{Bijnens1998}.}
   \footnotetext[4]{The intensity has been obtained after subtracting the contribution from the annihilation peak. A proper background subtraction has been taken into account.}
   \footnotetext[5]{Spin and parity are assigned on the basis of $R_{DCO}$ and $\Delta_{IPDCO}$ values for 526.4- and 385.9-keV transitions. This assignment is consistent with that reported in Ref.~\cite{Fant1990}.}
   \footnotetext[6]{$R_{DCO}$values are obtained using the gate on the 443.0 keV transition.}
    \footnotetext[7]{$R_{DCO}$values are obtained using the gate on the 138.4 keV transition.}
     \footnotetext[8]{$R_{DCO}$values are obtained using the gate on the 667.6 keV transition.}
     \footnotetext[9]{$R_{DCO}$ values are obtained using the gate on the 455.5 keV transition.}
      \footnotetext[10]{$R_{DCO}$ values are obtained using the gate on the 571.5 keV transition.}
      \footnotetext[11]{$R_{DCO}$ values are obtained using the gate on the 575.2 keV transition.}
\end{table*}
\begin{figure*}[ht!]
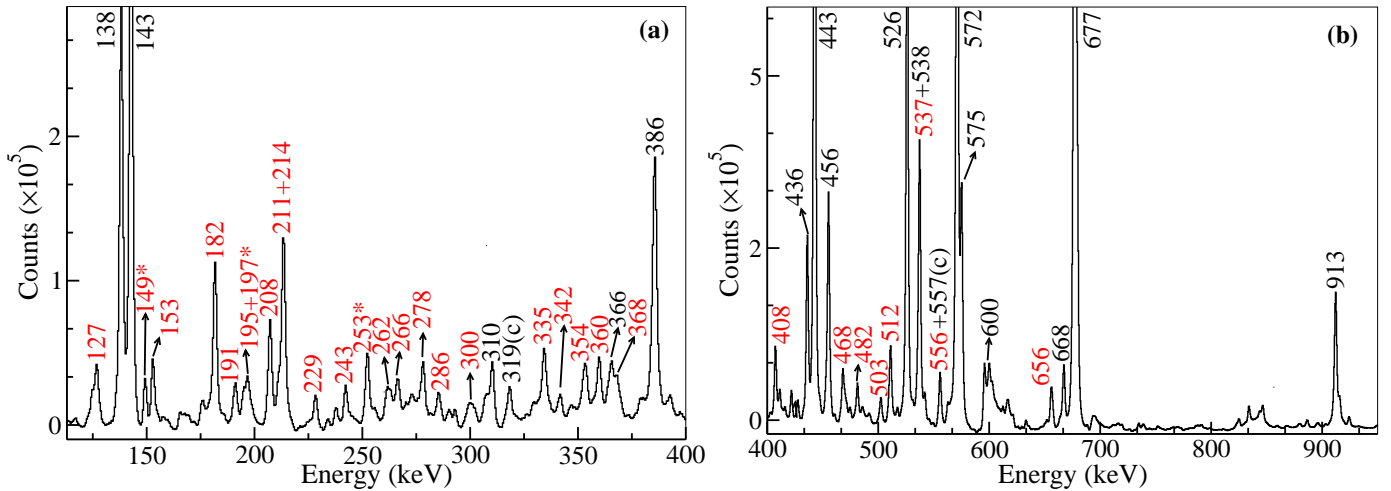

        \centering
        \begin{subfigure}[b]{0.52\textwidth}
                \centering
                \includegraphics*[width=\textwidth]{fig3a.eps}
                \label{fig1:GatesSpec1a}
        \end{subfigure}%
        \begin{subfigure}[b]{0.50\textwidth}
                \centering
                \includegraphics*[trim=0 0
 0 0,clip,width=\textwidth]{fig3b.eps} 
                \label{fig:GatesSpec1b}
        \end{subfigure}%
           \caption{Summed $\gamma$-ray spectra depicting transitions in coincidence with the 677.2-, 571.5-, and 443.0-keV $\gamma$ rays. The $\gamma$ rays with energy lower than 400 keV are shown in the left panel (a), while those with energies greater than 400 keV are shown in the right panel (b). The newly observed $\gamma$ rays are marked in red. The transitions marked with an asterisk are also from $^{202}$Po, however their position could not be ascertained in the level scheme. The 319- and 557-keV transitions marked with symbol (c) are contamination from $^{201}$Po.} 
        \label{fig:GatesSpec1}
\end{figure*}

In addition to the yrast sequence reported by Fant {\it et al.}~\cite{Fant1990}, several weak transitions were identified feeding the yrast $2^{+}$ and 4$^+$ states at 677.2 and 1248.7 keV, respectively. All these transitions were also reported in the $\beta$-decay study of $^{202}$At~\cite{Bijnens1998}, except for the 273.0-, 278.0-, and 411.9-keV transitions which are newly identified in the present study. These transitions are placed in the level scheme, as shown in Fig.~\ref{fig1:LS202po}, based on their $\gamma\gamma$ coincidence relationships. Further, the 617.1- and 619.2-keV transitions have been found to be dipole in nature, based on their $R_{DCO}$ values. The $\Delta_{IPDCO}$ value of the 617.1-keV $\gamma$ ray suggests its electric nature. Multipolarities of the remaining weak transitions feeding the yrast $2^{+}$ and 4$^+$ states could not be determined in the present work due to their weak intensities. Hence, the multipolarites of these transitions and the spin-parities of the corresponding excited states, wherever provided, have been adopted from the $\beta$-decay study~\cite{Bijnens1998}, except for the 479.4-keV transition. An $E2$ multipolarity was reported for the 479.4-keV transition based on the conversion coefficient measurements~\cite{Bijnens1998}, and hence $J^{\pi}$ = $(3^{-},7^{-})$ was tentatively assigned to the corresponding de-exciting level in Ref.~\cite{Bijnens1998}. The shell-model calculations in the present work, discussed in Sec.~\ref{discussion}, suggest $J^{\pi}=5^-$ as the lowest negative-parity state in $^{202}$Po. Therefore, we assign a tentative $J^{\pi}$ = $(7^{-})$ to the 2345.2-keV level which de-excites via the 479.4-keV $\gamma$ ray.

%

\subsection{Energy of the unobserved transition de-exciting the isomeric $8^+$ state}

\begin{figure}[hb!]
\includegraphics*[width=80mm]{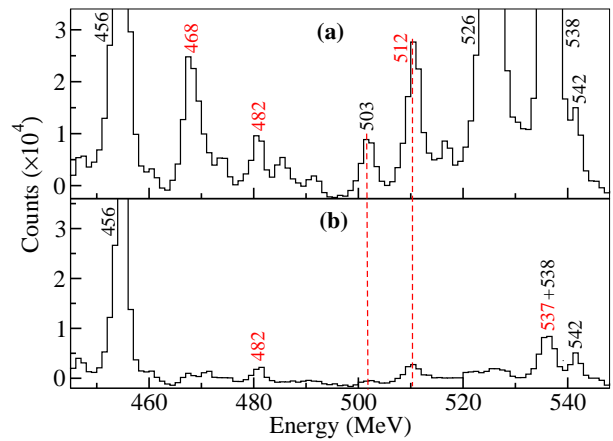}
\caption{\label{fig:502-511Spec} Gated spectra generated from the $\gamma-\gamma$ matrix, showing the $\gamma$ rays which are present in the coincidence gates of both the (a) 443.0-keV ($6^+ \rightarrow 4^+_1$) and 679.4-keV ($11^-_2 \rightarrow 9^-$) transitions, and (b) 443.0- ($6^+ \rightarrow 4^+_1$) and 912.5-keV ($11^-_1 \rightarrow 8^+$) transitions. The newly observed transitions are marked in red.}
\end{figure}

Prior to this work, the energy difference between the known isomeric $8^+$ state and the underneath yrast $6^+$ state was expected to be $<40$ keV~\cite{ensdf202Po}, though the $8^+\rightarrow6^+$ transition remained unobserved presumably due to complete internal conversion. In the present work, the energy of this unknown transition de-exciting the isomeric $8^+$ (114 ns~\cite{Sahab2024}) state to the $6^+$ state has been identified as 9.0(5) keV with the help of two parallel transitions, viz., 502.5 and 511.5 keV de-exciting the 2203.2-keV state to the $8^+$ and $6^+$ states, respectively. These two $\gamma$ rays were found in coincidence with the 677.2-, 571.5- and 443.0-keV transitions as can be seen in Fig.~\ref{fig:GatesSpec1}(b), but not observed in coincidence with the 526.4- and 912.5-keV transitions. Figure~\ref{fig:502-511Spec} shows the illustrative spectra depicting the common $\gamma$ rays in coincidence with the (a) 443.0- and 679.4-keV transitions, and (b) 443.0- and 912.5-keV transitions. The existence of the 502.5- and 511.5-keV transitions only in Fig.~\ref{fig:502-511Spec}(a) and their absence in Fig.~\ref{fig:502-511Spec}(b) further confirms their placement. Similar findings were reported for $^{200}$Po~\cite{Maj1990}, where the energy of the unobserved transition between the $8^+$ and $6^+$ states (12.2 keV) was identified with the help of parallel transitions.

A 502.8-keV $\gamma$-ray was also observed in Ref.~\cite{Bijnens1998}, establishing a state at $2194.3+\Delta$ keV, though the spin-parity of this state was tentatively assigned by them as $(7^-,8^-,9^-)$. The 511.5-keV $\gamma$ ray has been newly observed in the present work. The $R_{DCO}$ values for both the $\gamma$ rays suggest their dipole nature. Hence, $J^{\pi}$ = $7^-$ was assigned to the 2203.2-keV level on the basis of the measured $\Delta_{IPDCO}$ values for the two $\gamma$ rays.
The above $\gamma$ rays were also found in coincidence with all the sequences of transitions feeding the $9^-$ state at 2227.0 keV. 
These coincidence relationships suggest the presence of an unobserved 23.8-keV $\gamma$ ray between the isomeric $9^-$ ($<11$ ns~\cite{Fant1990}) and the newly established $7^-$ states at 2227.0 and 2203.2 keV, respectively.

\subsection{Newly observed medium-spin yrast and yrare states above the $8^+$ isomer}

The yet unobserved yrast $9^+$ and $10^+$ states have been established at $E_x=2616.5$ and 3044.2 keV, respectively, which de-excite to the $8^+$ isomeric state via $10^+\rightarrow9^+\rightarrow8^+$ decay path consisting of two newly observed $M1$ transitions of energies 427.7 and 915.9 keV. These two $\gamma$ rays were found in coincidence with the transitions of the yrast sequence below the $8^+$ state and parallel with the 526.4-keV ($9^-\rightarrow8^+$) and 912.5-keV ($11^-_1\rightarrow8^+$) transitions. The $10^+$ state at 3044.2 keV is further fed by a 127.0-keV dipole transition establishing an excited state at 3171.2 keV. Moreover, the placement of the 603.2-keV $\gamma$ ray above the yrast 8$^+$ state as reported earlier in Ref.~\cite{Bijnens1998}, has been confirmed on the basis of the coincidence relationships. The spin and parity of the corresponding de-exciting state at $E_x=2303.8$ keV has been adopted from Ref.~\cite{Bijnens1998}.

Along with the known sequence $13^-\rightarrow11^-_2\rightarrow9^-$ with $E2$ transitions of energies 537.6 and 679.4 keV, a parallel decay-path has been newly observed from the $13^-$ state to the $9^-$ state via a sequence of three $\gamma$-ray transitions of energies 153.2-, 407.7-, and 656.2 keV, classified as $M1$, $E2$ and $M1$ transitions respectively. The newly established yrast $12^-$ state is connected to the $11^-_1$ isomeric state (83 ns~\cite{Sahab2024}) through the 384.7-keV ($12^-\rightarrow11^-_2$) $M1$ transition, followed by 293.3-keV ($11^-_2\rightarrow11^-_1$) $\gamma$ ray. Moreover, a 600.5-keV $\gamma$ ray was observed in coincidence with the transitions de-exciting the 9$^-$ state at 2227.0 keV and placed parallel to the 656.2-keV ($10^-\rightarrow9^-$) transition as shown in Fig.~\ref{fig1:LS202po}. The coincidence relationship of this transition with the newly observed 153.2- and 407.7-keV transitions suggests an unobserved low-energy transition of 55.7 keV between the newly-established states at 2883.2- and 2827.5 keV. However, the ordering of the 600.5-keV and unobserved 55.5-keV transitions is not certain. Therefore, possibility of an alternative level at 2282.7 keV can not be discarded. Moreover, a direct feeding from the $13^-$ to the $11^-_1$ state via the 831.1-keV $E2$ transition has also been observed. 

A sequence of two transitions of energies 481.0 and 824.5 keV have also been newly observed in coincidence with the 526.4-keV $\gamma$-ray transition and in anti-coincidence with the 679.4- and 537.6-keV transitions. These two newly observed transitions have been placed in sequence above the $9^-$ isomeric state at 2227.0 keV. Though, the multipolatiry of these transitions, and hence, the spin-parity of the excited states at 2708.0- and 3532.5 keV could not be determined due to their low statistics. An additional $\gamma$-ray transition of energy 918.9 keV has also been found de-exciting the newly established 3532.5-keV state directly to the $11^-_1$ isomeric state, parallel to the 824.5-keV transition. The 918.9-keV transition was placed on the basis of its coincidence with the 385.9-keV ($11^-_1\rightarrow9^-$) transitions and all the lower-lying sequence of transitions de-exciting the $11^-_1$ state. Multipolarity of this transition could not determined due to low statistics.

A 180.7-keV $E1$ transition ($14^+\rightarrow13^-$) was reported earlier~\cite{Fant1990}, parallel to 575.5- and 138.4-keV transitions. However, the present data does not support these coincidence relations. In the present work, a strong 181.9-keV transition has been observed in coincidence with the transitions de-exciting the known 15$^{-}$ isomer (11 ns~\cite{Fant1990}) at 3582.4 keV (see Fig.~\ref{fig:GatesSpec3}(a)), but not in coincidence with those feeding the yrast 16$^+$ state. The $R_{DCO}$ and $\Delta_{IPDCO}$ values suggest an $M$1 nature for this transition. This newly observed transition has been placed directly above the 15$^{-}$ isomer based on its intensity and coincidence relationships with the newly observed transitions in the $M$1 sequence discussed in Sec.~\ref{results}(E).

\begin{figure}[b!]
\includegraphics*[width=80mm]{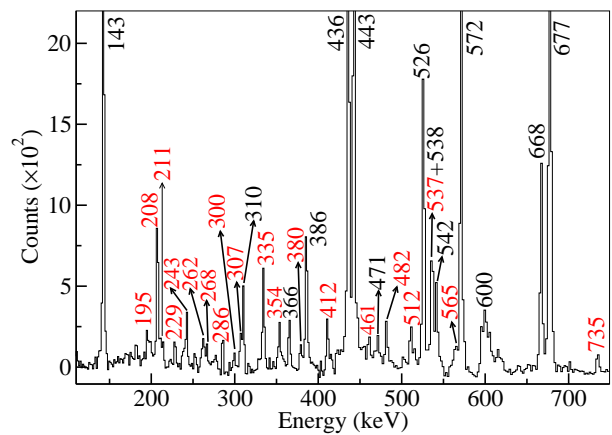}
\caption{\label{fig:GatesSpec2} A gated coincidence spectrum from $\gamma$-$\gamma$-$\gamma$ cube with a double-gate on 575.2- and 455.5-keV transitions. The newly observed transitions are marked in red.}
\end{figure}
\subsection{Newly observed $M1$ sequences above the yrast $16^+$ state}
\begin{figure}[hb!]
\includegraphics*[width=80mm]{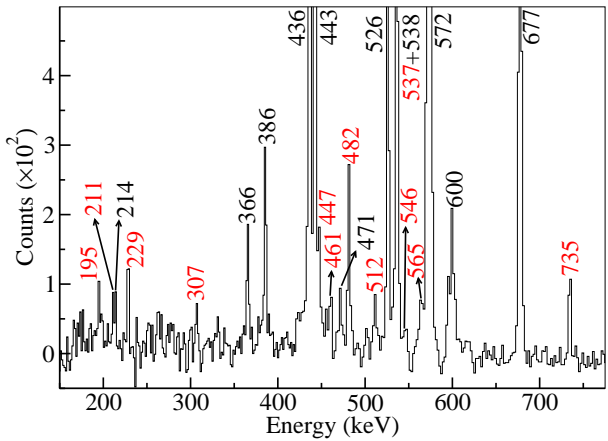}
\caption{\label{fig:455and668} A gated coincidence spectrum from $\gamma$-$\gamma$-$\gamma$ cube with a double-gate on 455.5- and 667.6-keV $\gamma$ rays highlighting the transitions in Group A of Fig. \ref{fig1:LS202po}. The newly observed transitions are marked in red.}
\end{figure}

\begin{figure}[t!]
\includegraphics*[width=80mm]{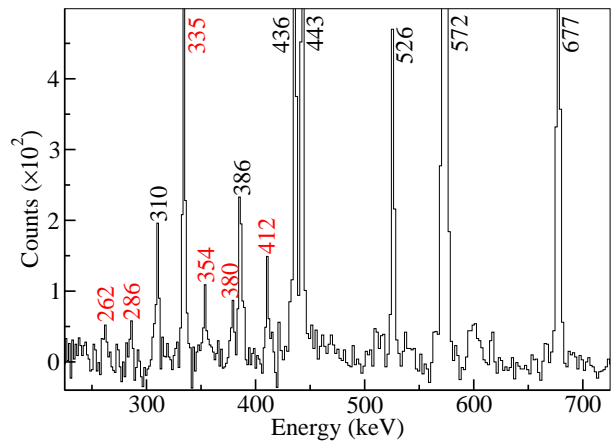}
\caption{\label{fig:455and207} A gated coincidence spectrum from $\gamma$-$\gamma$-$\gamma$ cube with a double-gate on 455.5- and 207.5-keV $\gamma$ rays highlighting some of the transitions in Group B of Fig. \ref{fig1:LS202po}. The newly observed transitions are marked in red.}
\end{figure}

\begin{figure}[hb!]
\includegraphics*[width=80mm]{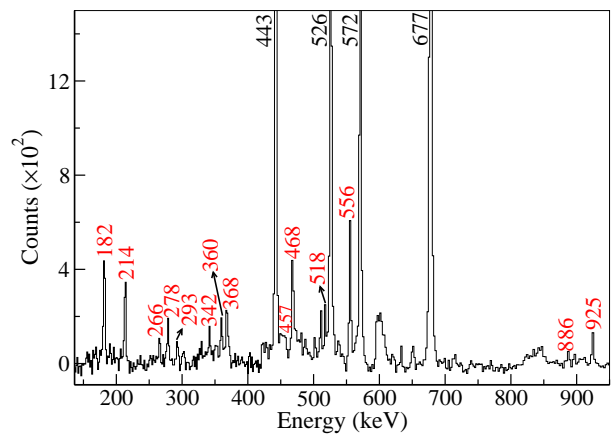}
\caption{\label{fig:GatesSpec3} $\gamma$-ray spectrum from $\gamma$-$\gamma$-$\gamma$ cube with a double gate on 537.6- and 138.4-keV transitions. The newly observed transitions are marked in red.}
\end{figure}

Previous to this work, only 9 transitions and 9 excited states were known above the $16^+_1$ state and the level scheme of $^{202}$Po was known up to a tentative spin of ($21^+$) and $E_x \simeq 5.5$ MeV~\cite{Fant1990}. These transitions/excited states have been revised in the present work with the assistance of coincidence analysis and the measured $R_{DCO}$ and $\Delta_{IPDCO}$ values, along with the addition of 24 new transitions above the $16^+_1$ state. This extends the level scheme up to an excitation energy of $\sim 8$ MeV and spin of 27$\hbar$. Figure~\ref{fig:GatesSpec2} shows a coincidence spectrum with a double-gate on the 455.5-keV ($16^+_1\rightarrow14^+$) and 575.2-keV ($14^+\rightarrow12^+$) transitions. The $\gamma$-ray transitions feeding the $16^+_1$ level have been primarily classified into two groups for the ease of discussion. Group A  contains the $\gamma$ rays feeding the $16^+_1$ level via the 667.6-keV ($18^+_1\rightarrow16^+_1$) transition (Fig.~\ref{fig:455and668}), while the transitions which are directly populating the $16^+_1$ level through three parallel transitions, viz., 542.2-, 207.5-, and 242.6-keV transitions are assigned as Group B. Figure~\ref{fig:455and207} shows those transitions of Group B which feed the $16^+_1$ state via the 207.5-keV transition. All the $\gamma$ rays in the above mentioned groups have been arranged according to their respective coincidences and relative intensities.



The yrast sequences of transitions in Group A, feeding the $18^+_1$ state at 4747.8 keV, which further de-excites to the $16^+_1$ state via the 667.6-keV $\gamma$-ray, mainly consists of dipole transitions, except for the 536.5- and 735.0-keV transitions, which were found to be quadrupole in nature. Prior to this work, the 667.6-keV transition was known to be decaying from the $18^+_3$ state at $4738.8+\Delta$ keV to the $16^+_1$ level~\cite{Fant1990}, whereas, in the present work, this transition was found to de-excite the yrast $18^+$ state at 4747.8 keV. A second 538-keV transition was reported in Ref.~\cite{Fant1990} as $18^+_1\rightarrow16^+_1$ transition, in addition to the $13^-\rightarrow11^-_2$ transition, and was placed parallel to the 667.6-keV $\gamma$-ray. The coincidence relationships in the present work do not support the above placement. In fact, the present data suggest two $\gamma$ rays with similar energies, viz. 536.5- and 537.8 keV in coincidence with each other and also in coincidence with the 667.6-keV $\gamma$ ray. Therefore, these transitions were placed above the $18^+_1$ state at 4747.8 keV, as shown in Fig.~\ref{fig1:LS202po}. An $E2$ and $M1$ multipolarities have been deduced for the 536.5- and 537.8-keV transitions, respectively, with the help of their $R_{DCO}$ and $\Delta_{IPDCO}$ values. These two transitions extend the yrast sequence up to $J^{\pi} = 21^+$ and excitation energy of 5822.1 keV.

Three mutually coincident $M1$ transitions of increasing energy, viz., 365.6, 471.3, and 565.3 keV have been placed above the 5822.1-keV level with the assistance of the coincidence relationships, extending the positive-parity yrast sequence up to an excited state of 7224.3 keV and $J^{\pi}=24^+$. Among these, the 565.3-keV $\gamma$-ray has been newly observed in our measurement, whereas the $\gamma$ rays similar to 365.6 and 471.3 keV were earlier reported by Fant {\it et al.}~\cite{Fant1990}. In their work, a $\gamma$-ray of 365.8 keV was reported as $(22^+)\rightarrow(20^+_1)$ $E2$ transition and another $\gamma$ ray of 470.4 keV was placed as $(21^+)\rightarrow(20^+_1)$ $M1$ transition, both being parallel to each other and feeding the state at $4823+\Delta$ keV~\cite{Fant1990}. The placement of the two transitions have been revised as shown in Fig.~\ref{fig1:LS202po}, based on the coincidence relationships. Two newly observed dipole transitions of energies 307.0 and 461.0 keV were identified and placed parallel to the 365.6-keV transition, as shown in Fig.~\ref{fig1:LS202po}. The 461.0-keV transitions have been found to have an electric nature as inferred from its $\Delta_{IPDCO}$ value, establishing the second excited $22^+$ state at 6283.1 keV. The yrast sequence has been further extended by a set of three dipole transitions of increasing energy, viz., 195.4, 211.3, and 267.5 keV, further extending the level scheme of $^{202}$Po up to an excitation energy of 7898.5 keV and spin 27$\hbar$. The multipolarity of these transitions could not be determined in the present work. Moreover, two mutually coincident dipole $\gamma$ rays of energies 213.9 and 546.2 keV were identified and placed parallel to the 565.3- and 195.4-keV transitions. A $(20^+_1)\rightarrow(18^+_1)$ $E2$ transition of energy 213.7 keV was reported in Ref.~\cite{Fant1990}, the placement and multipolarity of which are not consistent with the present study.

Parallel to the above mentioned yrast sequence feeding the $18^+_1$ state at 4747.8 keV, a 481.6-keV $M1$ transition was identified establishing the yrast $19^+$ state at 5229.4 keV. This state is further fed by a sequence of two transitions of energies 600.2 and 330 keV. The 600.2-keV $\gamma$-ray has been found to be quadrupole in nature whereas no multipolarity information could be obtained for the weak 330-keV transition due to low statistics. An $E2$ transition $(20^+_2)\rightarrow(18^+_1)$ of energy 600.4 keV was reported in Ref.~\cite{Fant1990} and placed on top of the ($18^+_1)\rightarrow16^+$, 538-keV transition, which was not observed in our work. Another sequence of two transitions of energies 735.0 and 447.0 keV has been newly observed in the present work feeding the $18^+_1$ state. The $R_{DCO}$ and $\Delta_{IPDCO}$ values of the 735.0-keV $\gamma$-ray suggest it to be $E2$ in nature, establishing the second excited $20^+$ state at 5482.8 keV. The 447.0-keV transition was found to be dipole in nature assigning a spin $J=21$ to the newly observed state at 5929.8 keV, though the parity of this state remains undetermined. An additional dipole $\gamma$-ray transition of energy 238.0 keV also directly feeds the $18^+_1$ state, parallel to the 481.6-, 536.5-, and 735-keV transitions, as can be seen in Fig.~\ref{fig1:LS202po}.

Group B consists of 11 $\gamma$-ray transitions above the yrast 16$^+$ state, forming one primary positive-parity sequence with several other parallel-decaying transitions. The positive-parity sequence commences with a 207.5-keV transition, observed as an unstretched $M1$ transition, followed by a cascade of stretched $M1$ transitions of energies 334.5, 310.4, 411.5, and 379.5 keV. These transitions were placed in the level scheme as shown in Fig.~\ref{fig1:LS202po}, based on their coincidence relationships and relative intensities. A dipole transition with energy 262.3 keV has been found feeding the 20$^+_3$ state at 5723.6 keV, though it's electric/magnetic nature could not be established. The $17^+_1$ state has been established at an excitation energy of 4622.2 keV, de-exciting to the $16^+_1$ state via three parallel paths. The spin and parity of this state is confirmed by the observed $M1$ nature of the 542.2-keV transition, which was earlier reported as a $18^+_2\rightarrow16^+$ E2 transition~\cite{Fant1990}. The second decay path of the 17$^+$ state consist of two consecutive dipole transitions, viz. 334.5 and 207.5 keV, which were identified as stretched and unstretched $M1$ transitions, respectively. The $\gamma$ rays forming the third parallel decay path, viz. 299.5 and 242.6 keV, are also identified as dipole and unstretched dipole, respectively, based on the corresponding $R_{DCO}$ values. Furthermore, Fant {\it et al.}~\cite{Fant1990} reported a $(19^+)\rightarrow18^+_2$, 310.8-keV transition, whereas, the corresponding 310.4-keV $\gamma$ ray observed in the present work is found to de-excite the $18^+_2$ state. The $18^+_3$ state has been also established at 4976.2 keV, depopulating via the 354.0-keV $M1$ transition to the $17^+_1$ state and by the 228.6-keV unstretched $M1$ transition to the $18^+_1$ state. This state is further populated by a dipole transition of energy 285.8 keV. Interestingly, no connection has been observed between the yrast $18^+$ and yrast $17^+$ states.

%
\subsection{Newly observed transitions and M1 sequence above the $15^-$ isomeric state}\label{M1_sequence}

 A total of 14 transitions have been newly identified in the present work above the $15^-$ isomeric state at 3582.4 keV, the highest negative-parity state known from the previous works~\cite{Beuscher1976, Fant1990}. These transitions are labeled as Group C in the level scheme (see Fig.~\ref{fig1:LS202po}) for the ease of discussion. Figure~\ref{fig:GatesSpec3} is an illustrative double-gated coincidence spectrum from $\gamma$-$\gamma$-$\gamma$ cube showing the $\gamma$ rays in coincidence with 537.6- and 138.4-keV transitions, where the newly identified transitions are marked in red color. A cascade of six $M1$ transitions of energies 181.9, 213.8, 468.4, 360.3, 278.4 and 266.4 keV has been observed feeding the $15^-$ isomeric state. These $\gamma$ rays have been found to be in mutual coincidence with each other, establishing a negative-parity yrast sequence of $M1$-decaying states up to an excitation energy of 5351.6 keV, as shown in Fig.~\ref{fig1:LS202po}. The $M1$ multipolarities of these transitions were confirmed from their measured $R_{DCO}$ and $\Delta_{IPDCO}$ values. This newly observed cascade of $M1$ transitions has been confirmed through the coincidence measurements and the placement of the transitions in the sequence has been made in decreasing order of their intensities. 

In addition to the $M1$ sequence, a few more transitions have also been discovered feeding the $15^-$ isomeric state at 3582.4 keV. A 556.1-keV $M$1 transition was found in coincidence with all the $\gamma$-ray transitions following the de-excitation of the $15^-$ isomeric state. This transition has been placed above the $15^-$ isomer and parallel to the newly observed $M1$ sequence, establishing the second excited $16^-_2$ state at 4138.5 keV. This state is further populated by three newly observed parallel $\gamma$-ray transitions of energies 342.0, 368.3 and 517.5 keV, all classified as $M1$ transitions, hence establishing the $17^-_2$, $17^-_3$, and $17^-_4$ states at 4480.5-, 4507.0-, and 4656.0 keV, respectively. An additional $\gamma$ ray of energy 457.0 keV has been found in coincidence with the 517.5- and 556.1-keV transitions and placed above the $17^-_4$ state, though the multipolarity of this transition remains undetermined. An additional $E2$ transition of 924.6 keV has been identified, which tends to directly de-excite the $17^-_3$ state to the $15^-$ isomeric state, parallel to the $17^-_3\rightarrow16^-_2\rightarrow15^-_1$ decay-path with the 368.3- and 556.1-keV $\gamma$ rays. Two more newly observed transitions, viz. 474.0 and 886.0 keV were identified, feeding the 3582.4-keV level. 

In addition to the transitions placed in the level scheme, there are a few more transitions, viz., 149, 197, and 253 keV [see Fig. 3(a)], which have been identified in the present work and originating from the excited states of $^{202}$Po. However, these transitions could not be placed in the level scheme.
\section{Theoretical Discussion}\label{discussion}

We have performed large-scale shell-model calculations to decipher the reported level scheme and the nucleonic configurations in $^{202}$Po. The PBPOP, a surface-delta interaction, is used with empirically fitted single-particle energies for the proton and neutron orbitals. All details about the fit can be followed in Ref.~\cite{Poppelier1988}. The results using this schematic interaction were found to be very similar to those with the Kuo-Herling matrix elements, and were shown in Ref.~\cite{Zwarts1985}. The PBPOP interaction comprises of seven proton orbitals $1d_{5/2}$, $1d_{3/2}$, $2s_{1/2}$, $0h_{11/2}$, $0h_{9/2}$,  $1f_{7/2}$, $0i_{13/2}$, with respective single-particle energies of $-9.86075$, $-8.05804$, $-8.007$, $-9.44850$, $-4.53318$, $-3.31374$, $-4.0$ MeV, and seven neutron orbitals $1f_{5/2}$, $2p_{3/2}$, $2p_{1/2}$, $0i_{13/2}$, $0i_{11/2}$, $1g_{9/2}$, $0j_{15/2}$ with respective single-particle energies of $-7.96499$, $-8.31831$, $-7.368$, $-9.09742$, $-3.42775$, $-4.11275$, $-4.0$ MeV. The calculations have been carried out with a natural proton shell closure at $Z=82$ and the neutron core in the interaction at $N=100$. No neutron-excitations above the neutron shell closure at $N=126$ have been considered. The active proton orbitals are hence $0h_{9/2}$, $1f_{7/2}$, $0i_{13/2}$ for the two valence protons and the 18 valence neutrons are active in $1f_{5/2}$, $2p_{3/2}$, $2p_{1/2}$, $0i_{13/2}$ orbitals. This valence space has been recently found to be sufficient for studying the level scheme and the electromagnetic properties of the Po isotopes in $A \approx 200$ region~\cite{Shukla2023}. The shell model Hamiltonian matrices are diagonalized by using the KSHELL code with the block Lanczos method~\cite{Shimizu2019}. The harmonic oscillator parameter is chosen as $41A^{-1/3}$. The center of mass correction is taken to be $\beta_{cm} = 100$. The shell model calculated results from PBPOP interaction demonstrate an overall agreement with the experimental data and with the results presented by S. Shukla {\it et al.}~\cite{Shukla2023}. To the best of our knowledge, this work discusses the first study for $^{202}$Po based on the PBPOP interaction, supporting the crucial role of pairing interaction in $A \approx 200$ region.

\begin{figure*}[ht!]
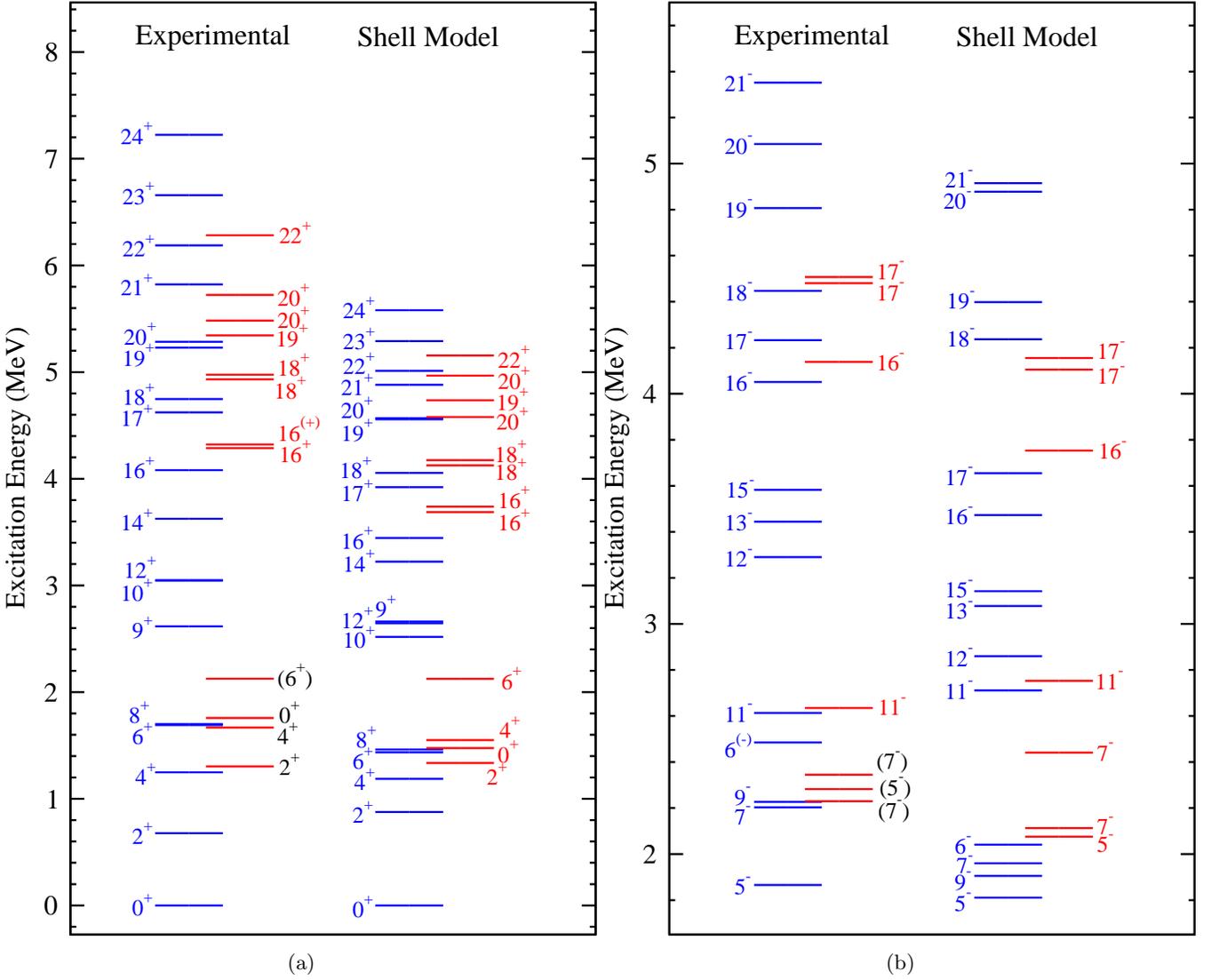

        \centering
        \begin{subfigure}[b]{0.50\textwidth}
                \centering
                \includegraphics*[width=\textwidth]{fig9a.eps}
                \caption{}
                \label{fig15a:AllPosSpins}
        \end{subfigure}%
        \begin{subfigure}[b]{0.50\textwidth}
                \centering
                \includegraphics*[trim=0 0
 0 0,clip,width=\textwidth]{fig9b.eps} 
                \caption{}
                \label{fig15b:AllNegSpins}
        \end{subfigure}%
        \caption{Comparison of the experimental and calculated level schemes of the $^{202}$Po nucleus, for the (a) positive parity states, and the (b) negative parity states. The yrast states are shown in blue color while all other excited states are shown in red color. The spin-parities of the states labelled in black color are adopted from Ref.~\cite{Bijnens1998}, except the second $0^+$ state for which both the energy and spin-parity are taken from Ref.~\cite{Bijnens1998}. All other experimental states which are plotted, are observed in the current study, unless otherwise stated.} 
        \label{fig:SM_Ex}
\end{figure*}

A comparison between the shell-model calculated and the observed level scheme for the positive parity states is presented in Fig.~\ref{fig:SM_Ex}(a).
The calculated results are shown only for the states for which measurements exist to compare. The yrast states are depicted in blue, while all other excited states are in red. The calculated results are found to be in overall good agreement with the experimental data, particularly for the yrast spin states up to the $8^+$ state. The calculated $E2$ energy gaps between the successive states ($2^+ \rightarrow 0^+$, $4^+ \rightarrow 2^+$, $6^+ \rightarrow 4^+$, and $8^+ \rightarrow 6^+$) decrease as is observed experimentally. Specifically, the relative gap between the $8^+$ and $6^+$ states is theoretically found to be 25 keV, which is in agreement to the experimentally determined energy gap of 9.0(5) keV. The shell model calculated $B(E2)$ values for the $8^+ \rightarrow 6^+$ transition is 133 $e^2 \text{fm}^4$, which is also in line with the measured value of 109(9) $e^2 \text{fm}^4$~\cite{Sahab2024}, supporting the presence of a proton $h_{9/2}$ dominated isomer. The calculations seemingly support the pure $h_{9/2}$ excitation scheme up to the $8^+$ state. The $8^+$ isomeric state and the $6^+$ state, to which the isomeric state de-excite via an $E2$ transition, are predominantly governed by two active protons in the $h_{9/2}$ orbital. The suggested configuration of the $8^+$ isomeric state from the calculations, is consistent with that suggested from the g-factor measurements~\cite{Hausser1976,Stone2020}.
Though, a significant configuration mixing is obtained for the yrast $2^+$ and $4^+$ states, in addition to the dominance of the $h_{9/2}$ protons. This can be noticed in the average occupancies of the active shell model orbitals, as listed in Table~\ref{tab:sm1}. For both the $6^+$ and $8^+$ states, the average occupancy of proton $h_{9/2}$ is close to two, with nearly negligible contributions from the neighboring $i_{13/2}$ proton orbital, and minimal contributions from the neutron side. In contrast to this, the wave function for the $0^+$, $2^+$, and $4^+$ states supports significant spread across the proton $h_{9/2}$ to proton $i_{13/2}$ and $f_{7/2}$ orbitals. This is accompanied by significant configuration mixing arising from the neutron $f_{5/2}$, $p_{3/2}$, $p_{1/2}$, and $i_{13/2}$ orbitals. To conclude, the $6^+$ and $8^+$ states are relatively pure proton $h_{9/2}^2$ states, whereas the lower spin states are found to have configuration mixing from the active proton and neutron orbitals.

\begin{table}[!htb]
\caption{\label{tab:sm1} Average occupancies of the active shell model proton and neutron orbitals for the first-excited states shown in Fig.~\ref{fig:SM_Ex}, for $^{202}$Po, using PBPOP interaction. The $<L_p>$ and $<L_n>$ list the magnitude of orbital angular momentum contributions from proton and neutron configurations, respectively. }
\begin{center}
\resizebox{0.49\textwidth}{!}{%
\begin{tabular}{| c | c c c | c c c c | c | c |}
\hline
\multicolumn{1}{|c|}{$^{202}$Po}&
\multicolumn{3}{c|}{Protons}&
\multicolumn{4}{c|}{Neutrons}& & \\
\hline
$J^\pi$ & $0h_{9/2}$ & $1f_{7/2}$ & $0i_{13/2}$ & $1f_{5/2}$ & $2p_{3/2}$ & $2p_{1/2}$ & $0i_{13/2}$ & $<Lp>$ & $<Ln>$\\
\hline
$0^+$ & 1.662  & 0.166 & 0.172 & 2.471  & 2.550  & 0.376  & 12.603 &  &  \\
$2^+$ & 1.834 & 0.087 & 0.079 & 2.468 & 2.335 & 0.498 & 12.699 & 1.119 & 0.971\\
$4^+$ & 1.822 & 0.089  & 0.090  & 2.426 & 2.417  & 0.327  & 12.829 & 2.122 & 2.049\\
$6^+$ & 1.977 & 0.020 & 0.003 & 2.482 & 2.556 & 0.356  & 12.607 & 6.297 & 0.219\\
$7^+$ & 1.884 & 0.113 & 0.002 & 2.366 & 2.252 & 0.386 & 12.996 & 6.808 & 0.695\\
$8^+$ & 1.938 & 0.059 & 0.003 & 2.494 & 2.540  & 0.391  & 12.575 & 8.408 &  0.249\\
$9^+$ & 1.935 & 0.062  & 0.003  & 2.360 & 2.210 & 0.484  & 12.946 & 8.023 & 1.632\\
$10^+$ & 1.932 & 0.065 & 0.002 & 2.344 & 2.220 & 0.464  & 12.973 & 8.415 & 2.211\\
$12^+$ & 1.734 & 0.150  & 0.116 & 3.012 & 3.023 & 0.420 & 11.546 & 0.120 & 11.019\\
$14^+$ & 1.849 & 0.090  &  0.061 & 2.865 & 2.881  & 0.609  & 11.646 & 1.088 & 12.174\\
$16^+$ & 1.802 & 0.113 & 0.085  & 3.003 & 2.951 & 0.365 & 11.680 & 1.454 & 13.945\\
$17^+$ & 1.860 & 0.124 & 0.015 & 2.964 & 3.029 & 0.427 & 11.580 & 6.369 & 10.316\\
$18^+$ & 1.946 & 0.034  & 0.020  & 3.024  & 2.952  & 0.388  & 11.636 & 4.868 & 12.745\\
$19^+$ & 1.919 & 0.067 & 0.014 & 3.006 & 3.074 & 0.404 & 11.516 & 7.602 & 11.158\\
$20^+$ & 1.944 & 0.049 & 0.008 & 2.832  & 3.161 & 0.257  & 11.750 & 5.718 & 14.310\\
$21^+$ & 1.878 & 0.116 & 0.005 & 2.833 & 3.125 & 0.298 & 11.744 & 6.483 & 14.505\\
$22^+$ & 1.979 & 0.018 & 0.002  & 2.750  & 3.295 & 0.215 & 11.740 & 6.981 & 15.14\\
$23^+$ & 1.965 & 0.028 & 0.007 & 2.631 & 3.383 & 0.254  & 11.731 & 7.962 & 15.307\\
$24^+$ & 1.984 & 0.015 & 0.001 & 2.604  & 3.439 & 0.221  & 11.735 & 8.509 & 15.816\\
\hline
$5^-$ & 1.693  & 0.157 & 0.150 & 2.644 & 2.656  & 0.317  & 12.383 & 0.021 & 4.874\\
$6^-$ & 1.694 & 0.156 & 0.150 & 2.413 & 2.849  & 0.334 & 12.404 & 0.025 & 5.564\\
$7^-$ & 1.686 & 0.159 & 0.155 & 2.534 & 2.784 & 0.363 & 12.319 & 0.084 & 6.854\\
$9^-$ & 1.687 & 0.159 & 0.155 & 2.664 & 2.750 & 0.309 & 12.278 & 0.176 & 8.688\\
$10^-$ & 1.803 & 0.094 & 0.104 & 2.513 & 2.498 & 0.485 & 12.504 & 1.062 & 8.715 \\
$11^-$ & 1.661 & 0.068 & 0.271 & 2.702 & 2.452  &0.340  & 12.505 & 3.729 & 7.237\\
$12^-$ & 1.770 & 0.116 & 0.113 & 2.503 & 2.671 & 0.255  & 12.571 & 1.347 & 10.412\\
$13^-$ & 1.953 & 0.039 & 0.008 & 2.568 & 2.783 & 0.357 & 12.293 & 5.628 & 7.704\\
$15^-$ & 1.948  & 0.046 & 0.007 & 2.572 & 2.819  & 0.299 & 12.311 & 7.191 & 8.273\\
$16^-$ & 1.876 & 0.122 & 0.002 & 2.526 & 2.820  & 0.355  & 12.299 & 7.883 & 8.555 \\
$17^-$ & 1.978 & 0.021 & 0.002 & 2.464 & 2.960  & 0.309 & 12.267 & 8.510 & 9.033 \\
$18^-$ & 1.958 & 0.041  & 0.001 & 2.496 & 2.650 & 0.163 & 12.691 & 7.163 & 11.083\\
$19^-$ & 1.762 & 0.139 & 0.099 & 3.427 & 3.374 & 0.363  & 10.836 & 0.364 & 17.823  \\
$20^-$ & 1.992 & 0.007 & 0.001  & 2.473  & 2.717  & 0.169  & 12.641 & 8.689 & 11.653 \\
$21^-$ & 1.831  & 0.100 & 0.069 & 3.282 & 3.431 & 0.371  & 10.916 & 1.311 & 19.049  \\
\hline
\end{tabular}}
\end{center}
\end{table}

\begin{table}[!htb]
\caption{\label{tab:sm2} Average occupancies of the active shell model orbitals for the other-excited states shown in Fig.~\ref{fig:SM_Ex}, for $^{202}$Po, using PBPOP interaction. The $<L_p>$ and $<L_n>$ list the magnitude of orbital angular momentum contributions from proton and neutron configurations, respectively. }
\begin{center}
\resizebox{0.49\textwidth}{!}{%
\begin{tabular}{| c | c c c | c c c c | c | c |}
\hline
\multicolumn{1}{|c|}{$^{202}$Po}&
\multicolumn{3}{c|}{Protons}&
\multicolumn{4}{c|}{Neutrons} & & \\
\hline
$J^\pi$ & $0h_{9/2}$ & $1f_{7/2}$ & $0i_{13/2}$ & $1f_{5/2}$ & $2p_{3/2}$ & $2p_{1/2}$ & $0i_{13/2}$ & $<Lp>$ & $<Ln>$\\
\hline
$0_2^+$ & 1.628 & 0.165 & 0.208 & 1.514 & 2.996 & 0.234 & 13.256 &  &    \\
$2_2^+$ & 1.752 & 0.118 & 0.130 & 2.250 & 2.359 & 0.381 & 13.009 & 0.688 & 1.332\\
$2_3^+$ & 1.653 & 0.158 & 0.189 & 2.141 & 2.491 & 0.275 & 13.094 & 0.291 & 1.916  \\
$4_2^+$ & 1.768 & 0.108 & 0.124 & 2.327 & 2.468 & 0.315 & 12.891 & 2.020 & 2.088\\
$4_3^+$ & 1.652 & 0.161 & 0.187 & 2.397 & 2.354 & 0.231 & 13.018 & 0.282 & 4.178   \\
$6_2^+$ & 1.753 & 0.115 & 0.132 & 2.366 & 2.053 & 0.216 & 13.365 & 1.392 & 4.690\\
$6_3^+$ & 1.857 & 0.122  & 0.021 & 2.307 & 2.244 & 0.414  & 13.035 & 7.583 & 2.003\\
$8_2^+$ & 1.190 & 0.802 & 0.008 & 2.511 & 2.484 & 0.414 & 12.591 & 7.589 & 0.453\\
${16}_2^+$ & 1.849 & 0.093 & 0.058 & 2.972 & 3.048 & 0.345 & 11.635 & 3.895 & 11.809\\
${16}_3^+$ & 1.784 & 0.121 & 0.095 & 3.097 & 2.856 & 0.318 & 11.729 & 1.892 & 13.569\\
${18}_2^+$ & 1.803 & 0.169 & 0.028 & 3.004 & 2.959 & 0.411 & 11.626 & 5.811 & 11.798\\
${18}_3^+$ & 1.805 & 0.147  & 0.048 & 3.068 & 2.884 & 0.323 & 11.725 & 4.253 & 13.296\\
${19}_2^+$ & 1.864 & 0.107 & 0.029 & 2.831 & 3.046 & 0.362 & 11.761 & 5.799 & 13.084\\
${20}_2^+$ & 1.958 & 0.039 & 0.003 & 3.001 & 3.106 & 0.386 & 11.507 & 7.951 & 11.858\\
${20}_3^+$ & 1.940 & 0.051 & 0.010 & 3.118 & 2.828 & 0.273 & 11.782 & 5.655 & 14.128\\
${22}_2^+$ & 1.874 & 0.121 & 0.005 & 2.753 & 3.189 & 0.299 & 11.759 & 7.155 & 14.902\\
\hline
$5_2^-$ & 1.689 & 0.158 & 0.152 & 2.490 & 2.805 & 0.318 & 12.387 & 0.006 & 4.523\\
$7_2^-$ & 1.695 & 0.156 & 0.150  & 2.549  & 2.664  & 0.375 & 12.412 & 0.131 & 6.478\\
$7_3^-$ & 1.723 & 0.142 & 0.135 & 2.357 & 2.568 & 0.783 & 12.292 & 0.085 & 6.445\\
${11}_2^-$ & 1.343 & 0.052 & 0.605 & 2.475 & 2.486 & 0.338  & 12.701 & 7.203 & 3.722\\
${16}_2^-$ & 1.954 & 0.045 & 0.002 & 2.547 & 2.796 & 0.307 & 12.349 & 8.245 & 8.141\\
${17}_2^-$ & 1.953 & 0.038  & 0.009 & 2.589 & 2.552 & 0.228 & 12.631 & 7.294 & 10.168   \\
${17}_3^-$ & 1.901 & 0.097 & 0.002 & 2.584 & 2.487 & 0.365 & 12.563 & 7.204 & 10.166\\
\hline
\end{tabular}}
\end{center}
\end{table}
%
There is an energy gap above the $8^+$ state in both the experimental and theoretical level schemes. The shell model reproduces the $9^+$, $10^+$, and $12^+$ states, although not in the same energy order as is observed experimentally. Notably, the measured energy of the yrast $10^+$ state almost coincides with that of the yrast $12^+$ state, whereas the relative gap between these states is overestimated in the calculated results. The shell model calculated $B(E2; 12^+ \rightarrow 10^+)$ is nearly zero. Experimentally, the $12^+$ state is not found to be de-exciting to the $10^+$ state via an $E2$ transition, which is consistent with the theoretical predictions. This can be understood in terms of the drastic change in the calculated wave functions, as can be noticed in the average occupancies for the ${10}^+$ and ${12}^+$ states listed in Table~\ref{tab:sm1}, leading to almost zero overlap between the initial and the final states using an $E2$ operator. The first-excited ${10}^+$ state is found to have more contributions from the proton orbitals though with non-zero configuration mixing from the neutron side. In contrast, the ${12}^+$ state is turned out to be an almost pure neutron state, with the dominating neutron $i_{13/2}$ orbital mixed with the neutron $f_{5/2}$, $p_{3/2}$, $p_{1/2}$ orbitals. This is also in-line with the generalized seniority interpretation~\cite{Maheshwari2021}. The calculated ${11}_1^-$ state is found to be mainly dominant by the neutron $i_{13/2}$ orbital mixed with neutron $f_{5/2}$, $p_{3/2}$, $p_{1/2}$ orbitals with non-zero configuration mixing from the proton side. The energy location for the ${11}_1^-$ state is overestimated in the calculations, as compared to the measured value. On the other hand, the calculated ${11}_2^-$ state is a proton dominating state, with $\pi(h_{9/2}i_{13/2})$ configuration. The calculated ${12}_1^+$ and ${11}_1^-$ states are both neutron dominating states, and consequently, the ${12}_1^+$ state was experimentally known to decay by $E1$ transition to the lower lying ${11}^-$ state as listed in Table~\ref{table:dco}, and was also found as an isomeric state~\cite{Fant1990}. In this way, the current shell model analysis has also explained the long-wondered missing $E2$ decay of the $\nu(i_{13/2})$ dominated ${12}^+$ isomer in $^{202}$Po isotope, especially when similar $E2$ decaying ${12}^+$ isomeric states exist in neighbouring Hg, Pb, and Po isotopic chains~\cite{Maheshwari2021}. Due to two extra protons in $h_{9/2},f_{7/2}$, and $i_{13/2}$ orbitals, the lower-lying $10^+$ state does not necessarily be a neutron dominating state in the $^{202}$Po isotope unlike in the neighboring Hg and Pb isotopes, where this state is predominantly neutron-dominated. The calculated ${10}_1^-$ and ${11}_2^-$ states lie at a lower excitation energy as compared to the ${12}_1^+$ state which was also the case in the experimental observation.
It is rather difficult to evaluate $E1$ and $E3$ transitions in the current calculations. We, therefore, could not investigate the $E3$ decaying $11^-$ isomer~\cite{Hausser1976,Stone2020}.
Although the calculated levels are significantly suppressed for higher spins above ${14}^+$, the shell model predicts three ${18}^+$ states clustered together, as observed experimentally. The calculations also suggest a cluster of three ${16}^+$ states at 3473, 3753, and 3997 keV, whereas the present experimental analysis established only two 16$^+$ states with firm spin-parity assignment. A comparison of these states with the shell-model results is shown in Fig.~\ref{fig:SM_Ex}(a). A closely spaced third excited state with $J=16$ has also been identified experimentally at $E_x$ = 4322.8 keV. However, its parity could not be deduced. The predicted cluster of three 16$^+$ states suggests a possible positive parity for the observed level at  $E_x$ = 4322.8 keV.

The shell model predicts the second $0^+$ state just above the yrast $8^+$ isomeric state, consistent with the experimental observation~\cite{Bijnens1998}. Since the $0^+_2\rightarrow0^+_1$ transition is fully converted, the $0_2^+$ state has not been observed in the present work, and has been adopted from Ref.~\cite{Bijnens1998}, marked with black labels in Fig.~\ref{fig:SM_Ex}. On comparing the average occupancies of the ground $0^+$ and second $0_2^+$ states, a clear rearrangement of neutron-pairs can be noticed from the $f_{5/2}$ to $(p_{3/2}$ $i_{13/2})$ orbitals for the $0_2^+$, as can be noticed in the average occupancies for the $0_2^+$ state listed in Table~\ref{tab:sm2}. The calculated locations of the second $2_2^+$ and $4_2^+$ states~\cite{Bijnens1998} also nearly reproduce the observed ones, though the calculated second excited $4_2^+$ state lies above the yrast $8^+$ isomeric state. 

\begin{table}[!htb]
\caption{\label{table:allE2} A list of shell model predicted $B(E2)$ values for all the observed $E2$ transitions listed in Table~\ref{table:dco}. The measured $B(E2)$ values are known only for the transitions de-exciting the $8^+$ and $15^-$ isomeric states, which are 109(9) $e^2fm^4$~\cite{Sahab2024}, and 349(10) $e^2fm^4$~\cite{ensdf202Po}, respectively. The standard effective charges for protons and neutrons are used as 1.5 and 0.5, respectively. }
\begin{center}
\resizebox{0.35\textwidth}{!}{%
\begin{tabular}{|c|cc|c|}
\hline
\multicolumn{1}{|c|}{Transition}&
\multicolumn{2}{|c|}{$E_\gamma$ (keV)}&
\multicolumn{1}{|c|}{$B(E2)$ $e^2fm^4$}\\
\hline
\multicolumn{1}{|c}{$E2$}&
\multicolumn{1}{|c|}{Exp.}&
\multicolumn{1}{c|}{Cal.}&
\multicolumn{1}{|c|}{Cal.}\\
\hline
$2^+ \rightarrow 0^+$ & 677 & 876 &   403.9  \\
$4^+ \rightarrow 2^+$ & 572 & 310 &  332.3 \\
$6^+ \rightarrow 4^+$ & 443 & 250 &   205.5  \\
$8^+ \rightarrow 6^+$ & 9 & 25 &  133.5  \\
$11_1^- \rightarrow 9^-$ & 386 & 806 &  232.0   \\
$11_2^- \rightarrow 9^-$ & 679 & 848 & 101.4  \\
$12^- \rightarrow 10^-$ & 408 & 207 &  160.2 \\
$13^- \rightarrow 11^-$ & 538 & 367 &   137.0 \\
$15^- \rightarrow 13^-$ & 138 & 64 &  136.0  \\
$14^+ \rightarrow 12^+$ & 575 & 578 &  453.8   \\
$16^+ \rightarrow 14^+$ & 456 & 222 &  364.0   \\
$18^+ \rightarrow 16^+$ & 668 & 609 &   290.0   \\
$20^+ \rightarrow 18^+$ & 537 & 514 &   71.0  \\
\hline
\end{tabular}}
\end{center}
\end{table}

\begin{table}[!htb]
\caption{\label{tab:allM1} A list of shell model predicted $B(M1)$ values for all the observed $M1$ transitions listed in Table~\ref{table:dco}. The gyromagnetic ratios used for the calculations are $g_l=1, g_s=5.585$ for protons and $g_l=0, g_s=-3.826$ for neutrons.}
\begin{center}
\resizebox{0.3\textwidth}{!}{%
\begin{tabular}{|c|cc|c|}
\hline
\multicolumn{1}{|c}{Transition}
&
\multicolumn{2}{|c|}{$E_\gamma$ (keV)}&
\multicolumn{1}{|c|}{$B(M1)$ $\mu_N^2$} \\
\hline
\multicolumn{1}{|c}{$M$1}&\multicolumn{1}{|c|}{Exp.}&
\multicolumn{1}{c|}{Cal.} & \multicolumn{1}{c|}{Cal.}\\
\hline
$9^+ \rightarrow 8^+$ & 916 & 1200 & 103.5\\
$16_2^+ \rightarrow 16_1^+$ & 208 & 242 & 5.8\\
$16_3^+ \rightarrow 16_1^+$ & 243 & 51   & 57.0\\
$17^+ \rightarrow 16_2^+$ & 335 & 234   & 250.0\\
$17^+ \rightarrow 16_3^+$ & 300 & 182   & 95.0\\
$17^+ \rightarrow 16^+$ & 542 & 476   & 50.0\\
$18_2^+ \rightarrow 17^+$ & 310 & 205   & 432.0\\
$18_3^+ \rightarrow 17^+$ & 354 & 253  & 71.0\\
$18_3^+ \rightarrow 18_1^+$ & 229 & 120   & 116.0\\
$19^+ \rightarrow 18^+$ & 482 & 252  & 50.0\\
$19_2^+ \rightarrow 18_2^+$ & 412 & 497   & 16.0\\
$21^+ \rightarrow 20^+$ & 538 & 314   & 300.0\\
$20_3^+ \rightarrow 19_2^+$ & 380 & 142  & 12.0\\
$22^+ \rightarrow 21^+$ & 366 & 155   & 155.0\\
$22_2^+ \rightarrow 21^+$ & 461 & 275   & 254.0\\
$23^+ \rightarrow 22^+$ & 471 & 277   & 148.0\\
$24^+ \rightarrow 23^+$ & 565 & 566   & 24.0\\
$6^- \rightarrow 5^-$ & 619 & 229   & 36.0\\
$10^- \rightarrow 9^-$ & 656 & 248  & 261.2\\
$12^- \rightarrow 11_2^-$ & 385 & 107  & 14.6\\
$13^- \rightarrow 12^-$ & 153 & 217   & 0.1\\
$16^- \rightarrow 15^-$ & 182 & 331   & 292.0\\
$17^- \rightarrow 16^-$ & 214 & 183   & 177.0\\
$18^- \rightarrow 17^-$ & 468 & 582   & 0.0\\
$19^- \rightarrow 18^-$ & 360 & 161   & 0.0\\
$20^- \rightarrow 19^-$ & 278 & 480   & 0.0\\
$21^- \rightarrow 20^-$ & 266 & 37   & 0.0\\
$16_2^- \rightarrow 15^-$ & 556 & 611   & 0.4\\
$17_2^- \rightarrow 16_2^-$ & 342 & 353   & 1.7\\
$17_3^- \rightarrow 16_2^-$ & 368 & 402   & 8.9\\
$17_4^- \rightarrow 16_2^-$ & 518 & 527   & 1.8\\
\hline
\end{tabular}}
\end{center}
\end{table}
\begin{figure*}[ht!]
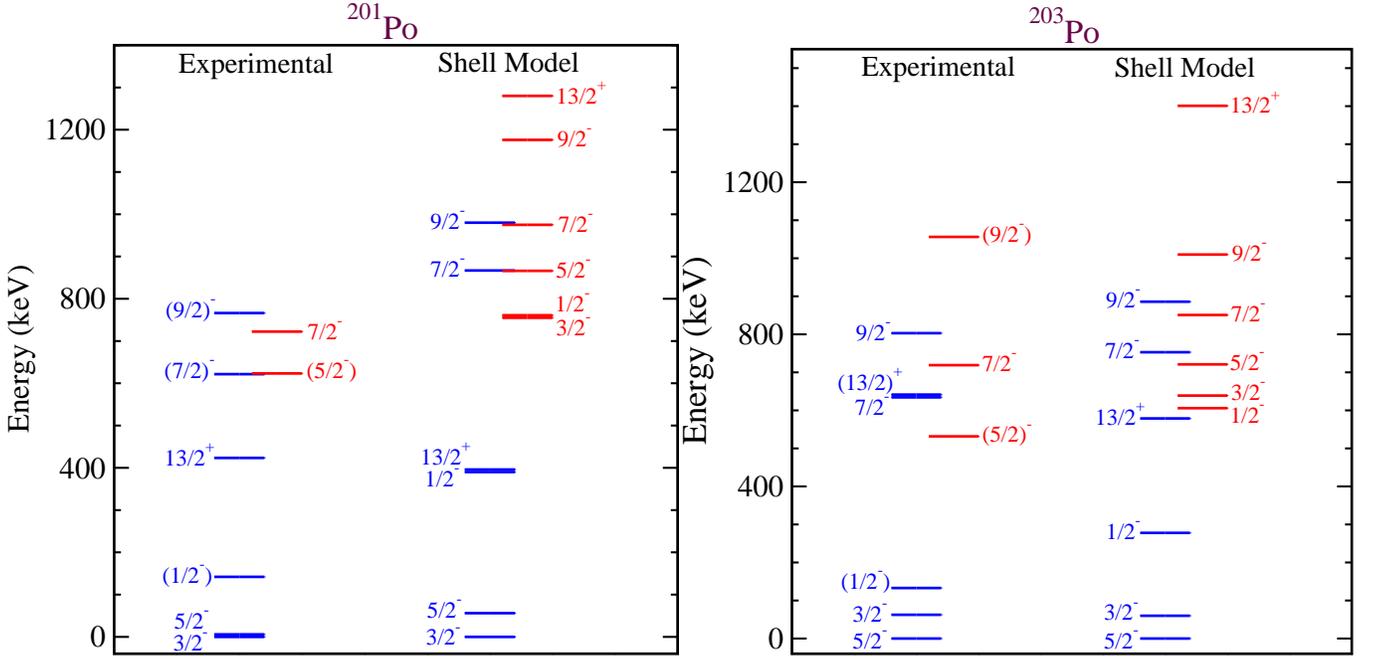

        \centering
        \begin{subfigure}[b]{0.50\textwidth}
                \centering
                \includegraphics*[width=\textwidth]{fig10a.eps}
                \label{fig7a:201Po}
        \end{subfigure}%
        \begin{subfigure}[b]{0.50\textwidth}
                \centering
                \includegraphics*[trim=0 0
 0 0,clip,width=\textwidth]{fig10b.eps} 
                \label{fig7b:203Po}
        \end{subfigure}%
           \caption{A comparison of experimental and calculated first and second-excited ${1/2}^-$, ${3/2}^-$, ${5/2}^-$, ${7/2}^-$, ${9/2}^-$ and ${13/2}^+$ states, in (a) $^{201}$Po, and (b) $^{203}$Po. The yrast states are shown in blue color, while the yrare states are shown in red color. The experimental data have been taken from Ref.~\cite{201Po} and~\cite{203Po} for $^{201}$Po and $^{203}$Po, respectively.} 
        \label{fig:Odd_A_Plot}
\end{figure*}

The first calculated negative parity state is the $5_1^-$ state, located at 1.81 MeV, which reasonably agrees with the experimental data, as shown in Fig.~\ref{fig:SM_Ex}(b). This state is found to be of nearly pure neutron $(p_{3/2}f_{5/2} i_{13/2})$ configuration from the shell model calculation. However, the calculated $7_1^-$ and $9_1^-$ states are obtained having the same configuration but their locations are significantly underestimated compared to the experimentally observed levels, and their level ordering is also not in agreement with the measurement. The calculations also suggest a cluster of $7^-$ states, as tentatively proposed in the experimental level scheme, shown in Fig.~\ref{fig:SM_Ex}(b). The calculations suggest the 6$^{-}$ and 12$^{-}$ states as neutron dominated states, though both the states are underestimated in energy, as compared to the experimentally observed levels. Overall, the calculated negative parity levels agree with the experimental data, although they are slightly suppressed in energy, than the experimentally observed states. The ${15}^-$ isomeric state is also suppressed, leading to a calculated $B(E2; 15^- \rightarrow 13^-)$ value of 136.5 $e^2$ $fm^4$, which is of the order of 3 times smaller than the experimentally observed value of 349(10) $e^2fm^4$~\cite{Hausser1976,Fant1990}. Both the calculated first-excited ${15}^-$ and ${13}^-$ states are from mixed proton and neutron configurations. Similarly, the calculated first-excited ${16}^-$, ${17}^-$, ${18}^-$, and ${20}^-$ states have mixed proton and neutron configurations. On the other hand, the other negative parity states such as the ${12}^-$, ${19}^-$, ${21}^-$, and ${22}^-$, are mainly dominated by neutron configurations. 

Table~\ref{table:allE2} lists the $B(E2)$ values predicted from the shell model calculations for the observed E2 transitions. The measured $B(E2)$ values are known only for the transitions de-exciting the $8^+$ and ${15}^-$ isomeric states.
The calculated $B(E2)$ for the ${20}^+ \rightarrow {18}^+$ transition is very low, though the involved $\gamma$-energy is in the order of 500 keV. It would be interesting to elaborate the (non)isomeric character of the ${20}^+$ using the lifetime analysis. It may be noted that the calculated $B(M1)$ values, as listed in Table~\ref{tab:allM1}, for the ${21}^- \rightarrow {20}^-$, ${20}^- \rightarrow {19}^-$, ${19}^- \rightarrow {18}^-$, and ${18}^- \rightarrow {17}^-$ transitions are found to be nearly zero, in contrast to the experimental observation of $M1$ connections. Though, there are other transitions predicted with strong $B(M1)$ values, such as $21^+ \rightarrow 20^+$, $17^- \rightarrow 16^-$, and $16^- \rightarrow 15^-$ transitions, etc. 
The calculated $B(M1)$ values for the positive parity yrast sequence of transitions $24^+\rightarrow23^+\rightarrow22^+\rightarrow21^+$ have been found to decrease with increasing spin. The shell model calculations predict a mixed proton-neutron configuration with protons in the $h_{9/2}, f_{7/2}$ orbitals and neutrons dominated by the $i_{13/2}$ orbital for all these states suggesting them to be a part of a band structure. Lifetime measurements for the newly established excited states in the observed $M1$ sequences along with detailed collective model calculations would further explain the phenomena active in these sequences.

In short, the shell model calculations explain the experimental positive parity states better than the negative parity states in $^{202}$Po, and also highlight the crucial role of pairing in $A \approx$ 200 region owing to the success of a surface delta interaction despite the large configuration mixing available due to so many high-$j$ active orbitals. 

In order to understand the absence of neutron excitations above the $N=126$ shell closure as considered in the current work, we further performed the shell model calculations for the lower-lying states in the neighboring odd-A $^{201}$Po and $^{203}$Po isotopes using the PBPOP interaction. A comparison of the calculated and experimental level energies is presented in Fig.~\ref{fig:Odd_A_Plot} for the first and second excited ${1/2}^-$, ${3/2}^-$, ${5/2}^-$, ${7/2}^-$, ${9/2}^-$, and ${13/2}^+$ states. Though the second excited $1/2^-$, $3/2^-$, and $13/2^+$ states are not yet known experimentally for both $^{201,203}$Po and second 9/2$^-$ in $^{201}$Po, an overall agreement is obtained for both the odd-A Po isotopes, especially for the states corresponding to the active single-particle orbitals in the chosen valence space, strengthening the suitability of the chosen interaction for this mass region and for the particular case of $^{202}$Po. In both the odd-A Po isotopes, the calculated ${7/2}^-$, and ${9/2}^-$ states are overestimated in energies on comparing with the experimentally known locations. 
In the odd-$A$ Po isotopes, the lower-lying states are dominated from the odd-neutron side, as expected. The ground states of both $^{201}$Po and $^{203}$Po are reproduced by the calculations. In $^{201}$Po, the ground state is a $p_{3/2}$ dominating, ${3/2}^-$ state. It may be noted that the $p_{3/2}$ orbital gets far from the Fermi surface as soon as one crosses $N=118$, $^{202}$Po, leading to the activeness of $f_{5/2}$ orbital, resulting ${5/2}^-$ as the ground state of $^{203}$Po. The neutron $p_{3/2}$, $f_{5/2}$ and $p_{1/2}$ orbitals get active altogether, leading to a spread in the resulting wave functions. Since these three neutron orbitals are low-$j$ orbitals, the main governance of the neutron $i_{13/2}$ orbital can be witnessed. This further competes with the proton $h_{9/2}$ orbital in even-even nuclei, where proton excitations are also possible, resulting in the complicated spectra such as the current case of $^{202}$Po. Such results in the neighboring odd-$A$ nuclei also answer about the neutron dominance of higher spins in $^{202}$Po shell model calculations. This analysis also overall supports the assumption of not considering the neutron-excitations above the $N=126$ shell closure for the lower-lying states in $^{202}$Po, though the suppression of energies especially at higher-spins may require some inclusion of core-excitations, which is beyond the scope of our current computational capacities. 

\section{Summary and conclusions}

The high-spin structures in the transition point nucleus $^{202}$Po have been investigated and an extended level scheme has been reported up to 27$\hbar$ of angular momentum and $\sim$8 MeV of excitation energy, with the addition of 57 newly observed $\gamma$-ray transitions extending the level scheme both horizontally and vertically. 
The energy of the unobserved transition de-exciting the isomeric $8^+$ state to the yrast $6^+$ state has been identified as 9.0(5) keV via two parallel transitions of energies 502.5 and 511.5 keV, feeding the $8^+_1$ and $6^+_1$ states, respectively.
One negative and two positive parity $M1$ sequences have been established above the $15^-_1$ and $16^+_1$ states, respectively, which need further experimental and theoretical investigation. The transitions reported from the previous works by Fant {\it et al.}~\cite{Fant1990} and Bijnens {\it et al.}~\cite{Bijnens1998}, have been mostly confirmed for the states with excitation energy below the $16^+_1$ state. The earlier known transitions above the $16^+_1$ state~\cite{Fant1990} have been revised on the basis of $\gamma-\gamma$ coincidence relationships and the measured $R_{DCO}$ and $\Delta_{IPDCO}$ values for the $\gamma$ rays, along with the addition of new transitions. In the negative parity region, the level scheme was earlier known up to $15^-$ isomeric state~\cite{Fant1990}, which has now been extended up to $J^{\pi} = 21^-$. 

The large-scale shell-model calculations were performed to understand the measured level scheme and configurations in $^{202}$Po. Though the first-excited $8^+$ isomeric state is found to be generated from nearly pure $h_{9/2}$ orbital, a significant amount of configuration mixing from both proton and neutron orbitals is responsible for the overall high-spin structure of this nuclide for both the positive and negative parity spins. The configuration mixing is found to play a crucial role in deciding the transition overlaps, ultimately influencing the half-lives and decays of the excited states. A particular case of the neutron-dominated ${12}^+$ isomeric state is hence discussed whose decay to the lower lying ${10}^+$ state is hindered due to the drastic change in the nucleonic configurations.
The used pairing governed PBPOP interaction is found to be sufficient in explaining the overall structure of $^{202}$Po, with $N=118$, 8 neutrons away from the $N=126$ shell closure. This suggests that the single-particle excitations with configuration mixing could dominate the complicated energy spectra, though the collective excitations start to appear as soon as one crosses ${16}^+$ and ${15}^-$ states, respectively. The shell model energies, though reproduced the higher spins and their ordering, are quite underestimated in energy as compare to the measured data. It may be due to missing core-excitations from both proton $Z=82$ and $N=126$ shell closures, though an overall agreement in $^{201,203}$Po isotopes hints towards the sufficiency of the chosen interaction, especially for the neutron orbitals below and above $N=118$.  

The collective picture for the high-spin states and the observed $M1$ sequences $\ge$ ${16}^-$, or ${17}^+$ along with the expected isomeric states need further investigation via lifetime measurements.
\begin{acknowledgments}
The authors thank the staff at TIFR-BARC Pelletron LINAC Facility for the smooth operation of the accelerator during the experiment. We acknowledge the financial support from the SERB-DST India under CRG (CRG/2022/005439). SS acknowledges the fellowship support from the Ministry of Human Resource Development, Government of India. BM gratefully acknowledges the financial support from the Croatian Science Foundation and the \'Ecole Polytechnique F\'ed\'erale de Lausanne, under the project TTP-2018-07-3554, with funds of the Croatian-Swiss Research Programme. This work is also supported by the Department of Atomic Energy, Government of India (Project Identification No. RTI 4002), and the Department of Science and Technology, Government of India (Grant No. IR/S2/PF-03/2003-II).

\end{acknowledgments}
%

\end{document}